\documentclass[reprint,amssymb, amsmath, aps, superscriptaddress, showpacs, footinbib, prb]{revtex4-1}

\usepackage{graphicx}
\usepackage{epsfig}
\usepackage{epstopdf}
\usepackage{subfigure}
\usepackage{tikz}

\usepackage{bm}
\usepackage{latexsym}
\usepackage{amssymb}
\usepackage{amsfonts}
\usepackage{amsmath}

\usepackage[colorlinks,bookmarks=false,citecolor=blue,linkcolor=red,urlcolor=blue]{hyperref}
\usepackage{verbatim}
\usepackage{wasysym}
\usepackage{booktabs}

\def\be{\begin{equation}}
\def\ee{\end{equation}}
\def\bea{\begin{eqnarray}}
\def\eea{\end{eqnarray}}

\def\vec{\mathbf}
\def\mc{\mathcal}
\def\bs{\boldsymbol}

\usepackage{url}

\usepackage{color}

\definecolor{darkblue}{rgb}{0,0.02,0.45}
\definecolor{darkred}{rgb}{0.45,0.02,0} 

\def\sysX{Cu$_3$Bi(SeO$_3)_2$O$_2$X}
\def\sysCl{Cu$_3$Bi(SeO$_3)_2$O$_2$Cl}
\def\sysBr{Cu$_3$Bi(SeO$_3)_2$O$_2$Br}

\newcommand{\eff}{\text{eff}}
\newcommand{\AFM}{\text{AFM}}

\newcommand{\rv}{\mathbf r}

\newcommand{\Dv}{\mathbf D}

\newcommand{\Lv}{\mathbf L}
\newcommand{\Sv}{\mathbf S}
\newcommand{\Gammav}{\mathbf{\Gamma}}

\usepackage{times}
\begin{document}

\title{Frustration and Dzyaloshinsky-Moriya anisotropy in the kagome francisites Cu$_3$Bi(SeO$_3)_2$O$_2$X}

\author{Ioannis Rousochatzakis}\email{rousocha@pks.mpg.de}
\affiliation{Max Planck Institut f$\ddot{u}r$ Physik komplexer Systeme, N\"othnitzer Str. 38, 01187 Dresden, Germany}

\author{Johannes Richter}
\author{Ronald Zinke}
\affiliation{Institute for Theoretical Physics, University of Magdeburg, P.O. Box 4120,
39016 Magdeburg, Germany}

\author{Alexander A. Tsirlin}
\email{altsirlin@gmail.com}
\affiliation{National Institute of Chemical Physics and Biophysics, 12618 Tallinn, Estonia}

\begin{abstract}
We investigate the antiferromagnetic canting instability of the spin-1/2 kagome ferromagnet, as realized in the layered cuprates \sysX\ (X=Br, Cl, and I). While the local canting can be explained in terms of competing exchange interactions, the direction of the ferrimagnetic order parameter fluctuates strongly even at short distances on account of frustration which gives rise to an infinite ground state degeneracy at the classical level. In analogy with the kagome antiferromagnet, the accidental degeneracy is fully lifted only by non-linear $1/S$ corrections, rendering the selected uniform canted phase very fragile even for spins-1/2, as shown explicitly by coupled-cluster calculations. To account for the observed ordering, we show that the minimal description of these systems must include the microscopic Dzyaloshinsky-Moriya interactions, which we obtain from density-functional band-structure calculations. The model explains all qualitative properties of the kagome francisites, including the detailed nature of the ground state and the anisotropic response under a magnetic field. The predicted magnon excitation spectrum and quantitative features of the magnetization process call for further experimental investigations of these compounds.
\end{abstract}


\maketitle

\section{Introduction}
Competing antiferromagnetic (AFM) interactions can prevent magnetic ordering down to very low temperatures and may even stabilize unconventional forms of ordering, such as the topological spin liquid phase originally envisaged by Anderson~\cite{Anderson1973}. One of the central models of highly frustrated magnetism that seems to realize this scenario in two spatial dimensions is the spin $S\!=\!1/2$ kagome antiferromagnet, which continues to attract enormous attention both in theory~\cite{HFMBook, Balents2010, Evenbly2010, Yan2011, CCMkagome2011, Depenbrock2012} and experiment~\cite{Shores2005, Mendels2007, Han2012, vanadate2013}. 
 
While most of the interest in the kagome lattice is associated with AFM interactions, here we show that frustration can have a dramatic impact even when the dominant exchange couplings are ferromagnetic (FM). This physics is realized in a series of layered cuprates \sysX\ (X=Br, Cl and I)~\cite{millet2001, pregelj2012, wang2012, miller2012}, where the kagome-like Cu$^{2+}$ layers are almost fully polarized on account of FM nearest-neighbor (NN) interactions. A directional, next-nearest-neighbor (NNN) AFM exchange $J_2$ gives rise to a local transverse AFM canting, but the lattice topology prevents any long-range ordering at the classical level, due to an infinite accidental ground state (GS) degeneracy. What is even more surprising is that the degeneracy remains infinite even at the quadratic spin-wave level, and is completely lifted only by non-linear $1/S$ corrections, in analogy with what happens in the kagome AFM~\cite{HKB1992, Chalker1992, RCC1993,Chubukov1992,CCMkagome2011}. As a result, the selected uniform canted phase is very fragile even for the extreme quantum-mechanical limit of $S\!=\!1/2$: The order-by-disorder stabilization energy is less than 0.5\% of the dominant NN ferromagnetic exchange,  as we show explicitly by coupled-cluster calculations. Ideally then, these systems would fluctuate over an infinite manifold of coplanar states (see below) down to very low temperatures, well below the experimental ordering temperatures $T_N\sim27$~K~\cite{millet2001, pregelj2012, wang2012, miller2012}.

To resolve this puzzle, we show that the minimal microscopic description of the kagome francisites must take into account the microscopic Dzyaloshinsky-Moriya (DM) interactions~\cite{dzyaloshinsky1958, moriya1960}. The latter are typically much weaker than the isotropic Heisenberg interactions, but here they play an important role due to frustration: The DM couplings lift the GS degeneracy and select the uniform canted phase already at the classical level, while the stabilization energy (linear in the DM couplings) can exceed significantly the corresponding energy gain by $1/S$ corrections in the isotropic case. 

The microscopic description presented below explains all qualitative properties observed experimentally, including the weak out-of-plane canting of the moments and the anisotropic response in a magnetic field. We also provide analytical results for a number of key quantities of experimental interest, and present the non-interacting spin-wave spectrum in the presence of DM anisotropy which can be used, in conjunction with future inelastic neutron scattering experiments, to extract accurate estimates of the microscopic parameters. Our study builds on extensive density-functional theory (DFT) calculations, analytical and numerical classical minimizations, semiclassical and coupled-cluster expansions, and exact diagonalizations (ED).

\begin{figure*}[!t]
\includegraphics{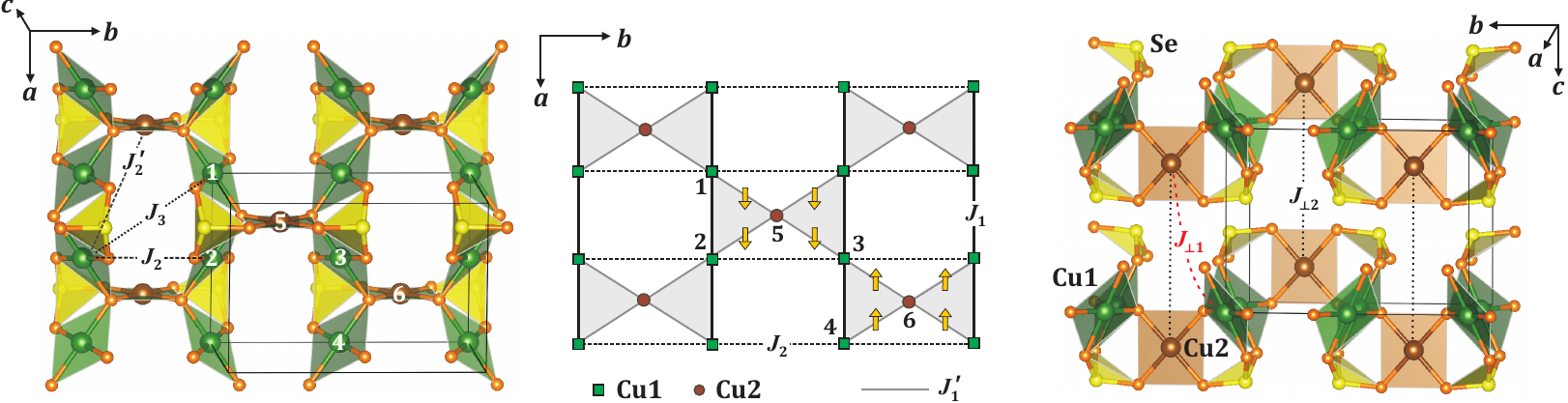}
\caption{\label{fig:structure}
(Color online) Left panel: kagome-like layers in the crystal structure of francisite. Green and brown colors denote the crystallographic positions Cu1 and Cu2, yellow triangles depict trigonal bipyramids SeO$_3$, the Bi and Cl atoms are not shown. Middle panel: projection of the spin lattice showing only the three leading in-plane couplings ($J_1$, $J_1'$, and $J_2$), with shaded triangles indicating the kagome units. The arrows depict the relative sign of the dominant, $\vec{a}$-component of the DM anisotropy on the $J_1'$-bonds. Right panel: stacking of the kagome layers and leading interplane couplings $J_{\perp1}$ and $J_{\perp2}$. The crystal structures are visualized using the \texttt{VESTA} software~\cite{vesta}.}
\end{figure*}

The three available kagome francisites (X=Br, Cl and I) are isostructural~\cite{millet2001}, and their main features are shown in Fig.~\ref{fig:structure}. These systems crystallize in the orthorhombic $Pmmn$ space group. The magnetic Cu$^{2+}$ kagome layers are perpendicular to the $\vec{c}$-axis and comprise two symmetry-inequivalent sites, denoted by Cu1 and Cu2. The former are inversion centers and the latter feature two orthogonal mirror planes, while there also exist glide plane symmetries involving non-primitive translations along $(\vec{a}\pm\vec{b})/2$, followed by a reflection in the $ab$-plane. 

The picture that emerges from a series of experiments~\cite{millet2001, pregelj2012, miller2012, wang2012} shows almost identical magnetic properties for the X = Br and Cl compounds. The long-range magnetic ordering sets in around $T_N\simeq 27.4$\,K and $24$\,K for the bromide and the chloride compound, respectively. Pregelj {\it et al.}~\cite{pregelj2012} have shown that each kagome plane displays a large net moment directed along $\vec{c}$, but these moments are canceled macroscopically because of the purely AFM interlayer coupling. In addition, single crystal magnetization measurements~\cite{pregelj2012,miller2012} point to a highly anisotropic response in a magnetic field, while at least two magnetic excitation modes have been observed at finite energies~\cite{wang2012,miller2012}. 

The leading (isotropic) exchange couplings have been already discussed by Pregelj {\it et al.}~\cite{pregelj2012}, and include two NN ferromagnetic couplings $J_1$ and $J_1'$, and the frustrating AFM coupling $J_2$ along the $\vec{b}$-axis, see Fig.~\ref{fig:structure}\,(b). The anisotropic DM couplings are finite on all bonds but, according to our DFT calculations, the DM vector on the $J_1'$ bonds is the most important ingredient and indeed, as shown below, it explains all qualitative properties of the francisites.

The article is organized as follows. First, we present our DFT band-structure calculations to extract the dominant isotropic and anisotropic interactions (Sec.~\ref{sec:DFT}). The theoretical analysis of the resulting spin model is then treated in two separate sections which focus, respectively, on the isotropic part of the model (\ref{sec:Hiso}) and the influence of the DM anisotropy (\ref{sec:HDM}). The former is organized into three separate subsections, which treat the GS manifold at the classical (\ref{sec:cgs1}) and the non-interacting spin-wave level (\ref{sec:lswt}), and the final non-linear order-by-disorder process, as found by coupled-cluster calculations (\ref{sec:GS_select_ccm}). Section~(\ref{sec:HDM}) is also organized in three subsections, which focus on the nature of the GS (\ref{sec:cgsdm}), the anisotropic response in a magnetic field (\ref{sec:MvsB}), and the non-interacting spin-wave spectrum (\ref{sec:LSWSpectraDM}). Our conclusions are given in Sec.~\ref{sec:discussion}, along with a qualitative comparison to reported experiments. Technical details and derivations are relegated in a series of Appendices.

\section{Microscopic magnetic model from DFT calculations}\label{sec:DFT}
\subsection{DFT methodology}
Isotropic and anisotropic magnetic couplings in francisites are obtained from density-functional (DFT) band-structure calculations performed in the full-potential local-orbital \texttt{FPLO} code~\cite{fplo} and the projector-augmented-wave \texttt{VASP} code~\cite{vasp1,*vasp2}. In both codes, the Perdew-Burke-Ernzerhof flavor of the exchange-correlation potential corresponding to the generalized gradient approximation (GGA) has been used~\cite{pbe96}. Strong correlations in the Cu $3d$ shell were treated on the mean-field level using the GGA+$U$ approach with the on-site Coulomb repulsion parameter $U_d=9.5$\,eV and Hund's exchange $J_d=1$\,eV~\cite{janson2012,lebernegg2013,nath2013}. 

For calculations, we used the crystallographic unit cell of francisite with 30 atoms, and 60-atom supercells doubled along either $b$ or $c$ directions. All calculations are done for the experimental orthorhombic $Pmmn$ crystal structures from Ref.~[\onlinecite{millet2001}]. Although several authors reported possible deviations from the $Pmmn$ symmetry at low temperatures~\cite{millet2001,pregelj2012}, no conclusive structural information is available in the literature. We tried to relax the francisite structure in several orthorhombic and monoclinic subgroups of $Pmmn$, but no appreciable energy gain and no significant deviations from the $Pmmn$ symmetry have been found.

Our procedure for the evaluation of the magnetic couplings is two-fold. On one hand, we analyze the GGA band structure without the GGA+$U$ correction and quantify the dispersions of relevant $d$ bands in terms of the tight-binding model. The resulting electron hoppings $t_i$ are a measure of AFM exchange couplings calculated as $J_i^{\AFM}\!=\!4t_i^2/U_{\eff}$, where $U_{\eff}$ is an effective on-site Coulomb repulsion. On the other hand, we evaluate exchange couplings $J_i$ and magnetic anisotropy parameters $\Dv_i$ and $\Gamma_i$ using total energies of spin configurations from GGA+$U$ via the four-configuration method introduced by Xiang~\textit{et al}~\cite{xiang2011}, and modified by our group to yield both DM vectors and symmetric anisotropy tensors $\Gammav_{ij}$~\cite{anisotropy}.  Altogether, we obtain parameters of the following spin Hamiltonian:
\be\label{eq:ham}
\mathcal H=\sum_{i<j} J_{ij}\,\Sv_i\cdot \Sv_j +\Dv_{i,j}\cdot \Sv_i\times\Sv_j+ \Sv_i\cdot\Gammav_{i,j}\cdot\Sv_j,
\ee
where $J_{ij}$ denote the isotropic Heisenberg exchange, $\Dv_{ij}$ are the DM vectors, and $\Gammav_{ij}$ are second-rank tensors that describe the symmetric portion of the anisotropic exchange.

\begin{figure*}[!t]
\includegraphics[width=0.88\textwidth]{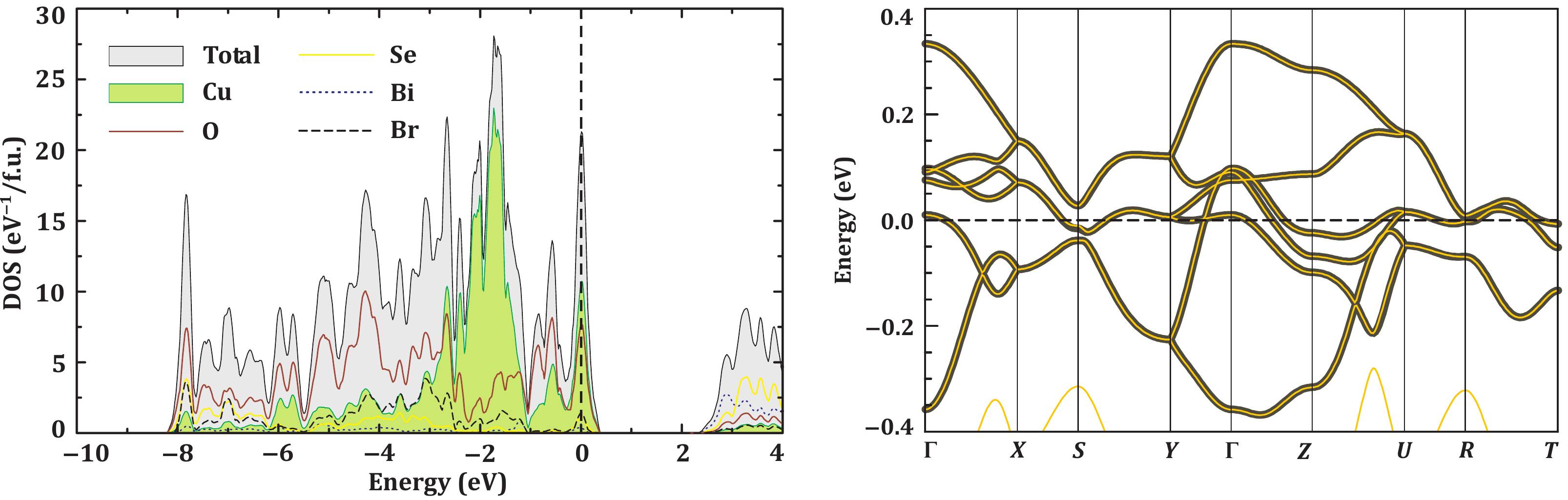}
\caption{(Color online) 
Left: GGA density of states (DOS) for \sysBr. The Fermi level is at zero energy. Right: GGA bands near the Fermi level (thin light lines) and their tight-binding fit using Wannier functions (thick dark lines). The notation of the $k$-points is as follows: $\Gamma(0,0,0)$, $X(\frac12,0,0)$, $S(\frac12,\frac12,0)$, $Y(0,\frac12,0)$, $Z(0,0,\frac12)$, $U(\frac12,0,\frac12)$, $R(\frac12,\frac12,\frac12)$, $T(0,\frac12,\frac12)$, all given in units of the reciprocal lattice parameters.
\label{fig:dos_and_bands}}
\end{figure*}

\subsection{Isotropic exchange}\label{sec:heis}
The electronic structure of francisites is typical for Cu$^{2+}$-based magnetic insulators. The GGA energy spectrum is metallic, because strong electronic correlations responsible for opening the band gap are missing in standard GGA. With GGA+$U$, we reproduce the insulating GS and obtain the energy gap of about $E_g=2.7$\,eV along with the magnetic moments of 0.88\,$\mu_B$ on the Cu atoms. Experimentally, the onset of electronic excitations in francisite is at 1.1\,eV according to optical spectra.\cite{miller2012} This discrepancy requires further investigation and may be due to impurity states facilitating electronic excitations at energies well below $E_g$.

The GGA energy spectrum of Fig.~\ref{fig:dos_and_bands}\,(left) shows Cu $3d$ bands in the vicinity of the Fermi level. These bands feature a large contribution of O $2p$ and a very small contribution of the halogen atoms, because these halogen atoms are only weakly bonded to Cu. The local environment of Cu$^{2+}$ is well understood as a CuO$_4$ plaquette with the Cu--O distances of $1.93-1.96$\,\r A (Fig.~\ref{fig:structure}). The halogen atoms X located above and below this plaquette form much longer Cu--X distances of $3.1-3.2$\,\r A. The reduction to the CuO$_4$ plaquette units is further justified by the electronic structure, where distinct crystal-field levels of Cu$^{2+}$ are observed. The states at the Fermi level are solely formed by the highest-lying Cu $d_{x^2-y^2}$ orbital, assuming that the local $x$ and $y$ axes are directed along the Cu--O bonds within the plaquette.

\begin{table*}
\caption{\label{tab:exchange}
Magnetic interactions in \sysX: Cu--Cu distances $d$ (in~\r A), hopping integrals $t_i$ (in~meV), AFM contributions to the exchange $J_i^{\AFM}$ (in~K) obtained from electron hoppings $t_i$ as $J_i^{\AFM}\!=\!4t_i^2/U_{\eff}$ with $U_{\eff}\!=\!4.5$\,eV,\cite{nath2014,lebernegg2014} and exchange couplings $J_i$ (in~K) obtained from GGA+$U$ calculations with $U_d\!=\!9.5$\,eV and $J_d=1$\,eV.}
\begin{ruledtabular}
\begin{tabular}{lcrrr@{\hspace{4em}}crrr}\medskip
             & \multicolumn{3}{c}{\sysCl} &                        & \multicolumn{4}{c}{\sysBr}                         \\
bond type            & $d_{\text{Cu--Cu}}$ & $t_{ij}$ & $J_{ij}^{\AFM}$ & $J_{ij}$  & $d_{\text{Cu--Cu}}$ & $t_{ij}$ & $J_{ij}^{\AFM}$ & $J_{ij}$ \\
\hline
$J_1$        &    3.177            & $-6$  &   0.4        & $-76$  &   3.195             &  $-5$ &    0.3       & $-75$ \\
$J_1'$       &    3.254            & $-33$ &   11         & $-66$  &   3.273             & $-30$ &      9       & $-67$ \\
$J_2$        &    4.818            & $-76$ &   60         & 55     &   4.847             & $-68$ &     48       & 49    \\
$J_2'$       &    5.548            & $-33$ &   11         & 2      &   5.579             & $-29$ &      9       & 1     \\
$J_3$        &    5.771            & $-28$ &    9         & $-3$   &   5.805             & $-28$ &      9       & $-3$  \\
$J_{\perp1}$ &    6.414            & $-11$ &    1         & $-0.4$ &   6.464             & $-10$ &      1       & $-0.3$\\
$J_{\perp2}$ &    7.233            &   18  &    3         & 2      &   7.287             &   16  &      3       &  1    \\
\end{tabular}
\end{ruledtabular}
\end{table*}

Six Cu atoms from two crystallographic positions in the unit cell (4 atoms from Cu1 and 2 atoms from Cu2, see Fig.~\ref{fig:structure}) form six $d_{x^2-y^2}$ bands, see Fig.~\ref{fig:dos_and_bands}\,(right). Their tight-binding analysis performed via the calculation of Wannier functions with the $d_{x^2-y^2}$ orbital character\cite{wannier} yields hopping parameters $t_i$ listed in Table~\ref{tab:exchange}. The resulting tight-binding model can be now extended to an effective one-orbital Hubbard model with the on-site Coulomb repulsion $U_{\eff}=4.5$\,eV.\cite{nath2014,lebernegg2014} In the strongly localized limit ($t_i\ll U_{\eff}$) and for low-lying excitations, this model is reduced to the Heisenberg model with AFM exchanges $J_i^{\AFM}=4t_i^2/U_{\eff}$. This way, we find relatively weak AFM contributions to the nearest-neighbor couplings $J_1$ and $J_1'$ and a more substantial AFM second-neighbor coupling $J_2$, in agreement with the preceding empirical analysis in Ref.~[\onlinecite{pregelj2012}]. Another second-neighbor coupling $J_2'$ and the third-neighbor coupling $J_3$ do not exceed 10\,K, whereas all other in-plane couplings are well below 1\,K. Regarding the interplane couplings, the largest interaction is between the Cu2 atoms (Fig.~\ref{fig:structure}, right).

Using GGA+$U$, we obtain accurate estimates of all relevant exchange couplings in francisites. The FM couplings $J_1$ and $J_1'$ form a weakly distorted kagome lattice frustrated by the second-neighbor AFM coupling $J_2$. Remarkably, these second-neighbor couplings are sparse and form only two bonds per atom for two thirds of the Cu atoms, only (Fig.~\ref{fig:structure}, middle). Their Cu1--O$\ldots$O--Cu1 superexchange pathways involve the short O$\ldots$O contact ($d_{\text{O}\ldots\text{O}}=2.62$\,\r A) on the edge of the Cu2O$_4$ plaquette and remind of the next-nearest-neighbor couplings in $J_1-J_2$ Cu$^{2+}$-based spin chains~\cite{[{For example: }][{}]drechsler2006}. Other in-plane couplings are much weaker, because they lack efficient Cu--O$\ldots$O--Cu pathways. 

The FM nature of $J_1$ and $J_1'$ conforms to the relatively low Cu--O--Cu bridging angles of $110.6^{\circ}$ and $113.8^{\circ}$, respectively. The Goodenough-Kanamori-Anderson rules suggest that the interaction should be ferromagnetic for bridging angles close to $90^{\circ}$. The FM--AFM crossover is expected for the bridging angles around $96-98^{\circ}$~\cite{braden1996}, yet in francisite the twisted geometry of the CuO$_4$ plaquettes may extend FM couplings to much higher bridging angles, as seen in the kagome mineral kapellasite\cite{fak2012} and in other Cu$^{2+}$-based quantum magnets.\cite{nath2013}

Remarkably, the isotropic couplings $J_i$ are nearly the same in the Cl and Br compounds. This similarity is rooted in their very similar crystal structures\cite{millet2001} and in the minor contribution of halogen $p$-states to the Cu $d_{x^2-y^2}$ bands at the Fermi level, see Fig.~\ref{fig:dos_and_bands}\,(left). The halogen atoms are located between the Cu--O planes. Their weak bonding to Cu$^{2+}$ implies minor influence of halogen on structural details and, therefore, on the magnetic couplings. In the following, we present calculated magnetic anisotropy only for the Cl compound. These results are directly applicable to the bromide compound as well.


\subsection{Magnetic anisotropy}\label{sec:anis}
Experimental studies suggest that the magnetic susceptibility and magnetization process of Cu$_3$Bi(SeO$_3)_2$O$_2$X are highly anisotropic.\cite{pregelj2012} To explore the origin of this anisotropy, we evaluate three different anisotropic terms in the spin Hamiltonian for the chloride compound. 

\subsubsection{DM anisotropy}
First, we analyze the DM couplings listed in Table~\ref{tab:dm}. Two main symmetry elements of the francisite structure are inversion centers at the Cu1 site and two orthogonal mirror planes passing through the Cu2 site. The symmetry of Cu2 leads to the following relations between the various components of the DM vectors on the $J_1'$ bonds (see Fig.~\ref{fig:structure}(b)):
\bea
&&\vec{D}_{5,1} 
\equiv \left(d_a',d_b',d_c'\right),~~
\vec{D}_{5,2} 
= \left(d_a',-d_b',-d_c'\right),\label{eq:DMsJ1pA} \\
&&\vec{D}_{5,3} 
= \left(-d_a',-d_b',d_c'\right),~~
\vec{D}_{6,4} 
= \left(-d_a',d_b',-d_c'\right)~.\label{eq:DMsJ1pB}
\eea
Additionally, the inversion symmetry at Cu1 sites yields $\Dv_{5,3}=\Dv_{6,3}$, $\Dv_{5,1}=\Dv_{6-\vec{a}-\vec{b},1}$, etc. 
For the weaker DM couplings on the $J_1$ and $J_2$ bonds, symmetry necessitates that:
\bea\label{eq:dms}
&&\vec{D}_{1,2} = \vec{D}_{1+\vec{a},2} \equiv \left(0,d_{1b},d_{1c}\right),~\nonumber\\
&&\vec{D}_{3,4} = \vec{D}_{3,4-\vec{a}} = \left(0,-d_{1b},d_{1c}\right)~,\nonumber\\
&&\vec{D}_{2,3} = \vec{D}_{2+\vec{b},3}\equiv \left(d_{2a},0,d_{2c}\right),~\nonumber\\
&&\vec{D}_{1,4-\vec{a}} = \vec{D}_{1+\vec{b},4-\vec{a}} = \left(d_{2a},0,-d_{2c}\right)~.
\eea
The {\it ab initio} values of the above parameters can be extracted from Table~\ref{tab:dm} (in K): $d_a'=12.1$, $d_b'=-4.7$, $d_c'=-4.4$, $d_{1b}=2.6$, $d_{1c}=2.2$, $d_{2a}=1.6$ and $d_{2c}=2.7$. 
So, the dominant DM vectors are the ones on the $J_1'$ bonds and in particular the component $d_a'$ along the $\vec{a}$-axis. The sign of this component is depicted by arrows in Fig.~\ref{fig:structure}\,(b) and, as shown in Eq.~(\ref{eq:DMsJ1pA}-\ref{eq:DMsJ1pB}), alternates from one crossed plaquette to the next, which is essential for the stabilization of the uniform canted phase. 

\begin{table}[!b]
\caption{\label{tab:dm}
DM couplings $\Dv_{i,j}$ (in~K) and relevant atomic positions $\rv_i,\rv_j$ given in crystallographic coordinates for Cu$_3$Bi(SeO$_3)_2$O$_2$Cl. The fractional component for the position of Cu2 (atom 5) is $z=0.208$~\cite{millet2001}.
}
\begin{ruledtabular}
\begin{tabular}{ccccc}
bond        & $\rv_i$      & $\rv_j$           & $\Dv_{ij}$          & $|\Dv_{i,j}|/|J_{ij}|$ \\
\hline
 $J_1$  & $\vec{r}_1\!=\!(0,0,0)$  		  &$\vec{r}_2\!=\!(\frac{1}{2},0,0)$& $(0,2.6,2.2)$    & 0.045           \\
 $J_1'$ &  $\vec{r}_5\!=\!(\frac{1}{4},\frac{1}{4},z)$&$\vec{r}_2\!=\!(\frac{1}{2},0,0)$ & $(12.1,4.7,4.4)$ & 0.208           \\
 $J_2$  & 	 $\vec{r}_2\!=\!(\frac{1}{2},0,0)$&$\vec{r}_3\!=\!(\frac{1}{2},\frac{1}{2},0)$ & $(1.6,0,2.7)$   	 & 0.057          
\end{tabular}
\end{ruledtabular}
\end{table}

\subsubsection{Symmetric portion of the exchange anisotropy}
Next, we consider the symmetric portion of the exchange anisotropy, described by the second-rank tensor $\bs{\Gamma}_{ij}$ of Eq.~(\ref{eq:ham}). For the three bonds specified in Table~\ref{tab:dm} (the tensors on all remaining bonds follow by symmetry), we find:
\begin{gather*}
\Gammav_{1,2}=\left(
\begin{array}{ccc}
  0.0 & 0.0 & 0.0 \\
  0.0 & 1.6 & 1.4 \\
  0.0 & 1.4 & 1.1 \\
\end{array}
\right),\quad
\Gammav_{5,2}=\left(
\begin{array}{ccc}
  1.4 & 0.9 & 0.6 \\
  0.9 & 0.0 & 0.5 \\
  0.6 & 0.5 & 0.0 \\
\end{array}
\right), \\
\Gammav_{3,2}=\left(
\begin{array}{ccc}
  0.3 & 0.0 & 0.0 \\
  0.0 & 0.0 & 0.0 \\
  0.0 & 0.0 & 0.2 \\
\end{array}
\right),
\end{gather*}
consistent with all available symmetries (Note that the component $\Gamma_{3,2}^{xz}$ does not vanish by symmetry, but is found to be well below 0.1\,K). The resulting numerical values of the symmetric anisotropy tensors $\bs{\Gamma}_{ij}$ are too small to play any appreciable role in the magnetism of the kagome francisites.

\subsubsection{$\vec{g}$-tensor anisotropy}\label{sec:gtensors}
For completeness, let us also discuss the anisotropy of the electronic $\vec{g}$-tensors, which stems from the relativistic spin-orbit coupling. In the absence of this coupling, the orbital angular momentum is completely quenched when the crystalline electric field is of sufficiently low symmetry~\cite{Fazekas}. The spin-orbit coupling restores a weak orbital moment by admixing a finite amplitude of excited orbital states into the single-ion GS, leading to an anisotropic correction $\delta\vec{g}$ to the spectroscopic $\vec{g}$-tensor. We can capture this correction in a fully relativistic framework by calculating the matrix $\mathbb L\!=\!(\Lv_a,\Lv_b,\Lv_c)$, formed by the orbital moments $\Lv_\vec{v}$ generated by fixed spin moments of length $1/2$ along $\vec{v}\!=\!\vec{a}$, $\vec{b}$, and $\vec{c}$, respectively.
With the total angular momentum given by $\vec{J}\!=\!\vec{L}\!+\!2\vec{S}$, we may replace the Zeeman energy in a magnetic field $-\mu_B \vec{J}\cdot \vec{H}$ with $-\mu_B \vec{S}\cdot\vec{g}\cdot\vec{H}$, where $\vec{g}_{\alpha\beta}\!=\!2\delta_{\alpha\beta}\!+\!(\delta\vec{g})_{\alpha\beta}$, and $\delta \vec{g}\!=\!2~\mathbb{L}$. 

By symmetry, the matrices $\mathbb{L}$ take the following form in the $\{\vec{a},\vec{b},\vec{c}\}$ reference frame for the Cu1 site 1 and the two Cu2 sites (see Fig.~\ref{fig:structure}):
\begin{gather*}
\mathbb{L}_{5}\!=\!\mathbb{L}_{6}\!=\!\!\left(\!\!\begin{array}{rrr}
b_1 & 0 & 0 \\ 
0 & b_2 & 0 \\ 
0 & 0 & b_3 
\end{array}\!\right)\!, ~~
\mathbb{L}_{1}\!=\!\!\left(\!\!\begin{array}{rrr}
a_{11} & a_{12} & a_{13} \\ 
a_{12} & a_{22} & a_{23} \\ 
a_{13} & a_{23} & a_{33} 
\end{array}\!\right).
\end{gather*}
The remaining Cu1 matrices, $\mathbb{L}_2$, $\mathbb{L}_3$, and $\mathbb{L}_4$, can be obtained from $\mathbb{L}_1$ by replacing, respectively, $(a_{12},a_{13})\!\to\!-(a_{12},a_{13})$, $(a_{13},a_{23})\!\to\!-(a_{13},a_{23})$, and $(a_{12},a_{23})\!\to\!-(a_{12},a_{23})$. 
We find:
$a_{11}\!=\!0.099$, $a_{12}\!=\!-0.055$, $a_{13}\!=\!0.028$, $a_{22}\!=\!0.141$, $a_{23}\!=\!-0.038$, $a_{33}\!=\!0.100$, and 
$b_1\!=\!0.198$, $b_2\!=\!0.066$, $b_3\!=\!0.066$.
For Cu2, the orbital correction is diagonal, as expected by the $mm2$ ($C_{2v}$) symmetry of the Cu2  crystallographic position. Furthermore, the largest correction is along $\vec{a}$, i.e., perpendicular to the respective CuO$_4$ plaquette, similar to other Cu$^{2+}$ oxides.\cite{nath2014} 
For Cu1, the eigenvalues of the $\mathbb L$ matrices are 0.200, 0.078, and 0.062, and the eigenvector corresponding to the largest eigenvalue is again vertical to the corresponding CuO$_4$ plaquette. So the orbital correction to the $\vec{g}$-tensor of the Cu1 sites closely resembles that of Cu2.

As we show below, the spins $\vec{S}_{1}$ and $\vec{S}_{2}$ are almost entirely parallel to each other and the same is true for the spins $\vec{S}_{3}$ and $\vec{S}_{4}$. This means that, for most purposes, we can replace $\mathbb{L}_{1}$ and $\mathbb{L}_{2}$ with $\mathbb{L}_{1,2}^{\text{av}}\!=\!(\mathbb{L}_1\!+\!\mathbb{L}_2)/2$,  and $\mathbb{L}_{3}$ and $\mathbb{L}_{4}$ with $\mathbb{L}_{3,4}^{\text{av}}\!=\!(\mathbb{L}_3\!+\!\mathbb{L}_4)/2$:
\be
\mathbb{L}_{1,2}^{\text{av}}\!=\!\!\left(\!\!\begin{array}{rrr}
a_{11} & 0 & 0 \\ 
0 & a_{22} & a_{23} \\ 
0 & a_{23} & a_{33} 
\end{array}\!\!\right)\!, 
\mathbb{L}_{3,4}^{\text{av}}\!=\!\!\left(\!\!\begin{array}{rrr}
a_{11} & 0 & 0 \\ 
0 & a_{22} & -a_{23} \\ 
0 & -a_{23} & a_{33} 
\end{array}\!\!\right). \label{eq:a23}
\ee
Effectively then, the $\vec{a}$-axis is one of the principal axes for Cu1 sites too, while the remaining principal axes of $\mathbb{L}_{1,2}^{\text{av}}$ and $\mathbb{L}_{3,4}^{\text{av}}$ are rotated with respect to each other (by $\sim\!61^\circ$) due to the different signs in front of $a_{23}$.  
This off-diagonal element couples an external magnetic field along the $\vec{b}$-axis to the AFM component $(\vec{S}_{1}\!-\!\vec{S}_3)^c$, or a field along the $\vec{c}$-axis to $(\vec{S}_{1}\!-\!\vec{S}_3)^b$. The latter component is particularly large in the zero-field GS due to the $J_2$ coupling, and so this weak coupling may give rise to observable field-induced effects.


\begin{figure*}[!t]
\includegraphics[width=0.9\linewidth]{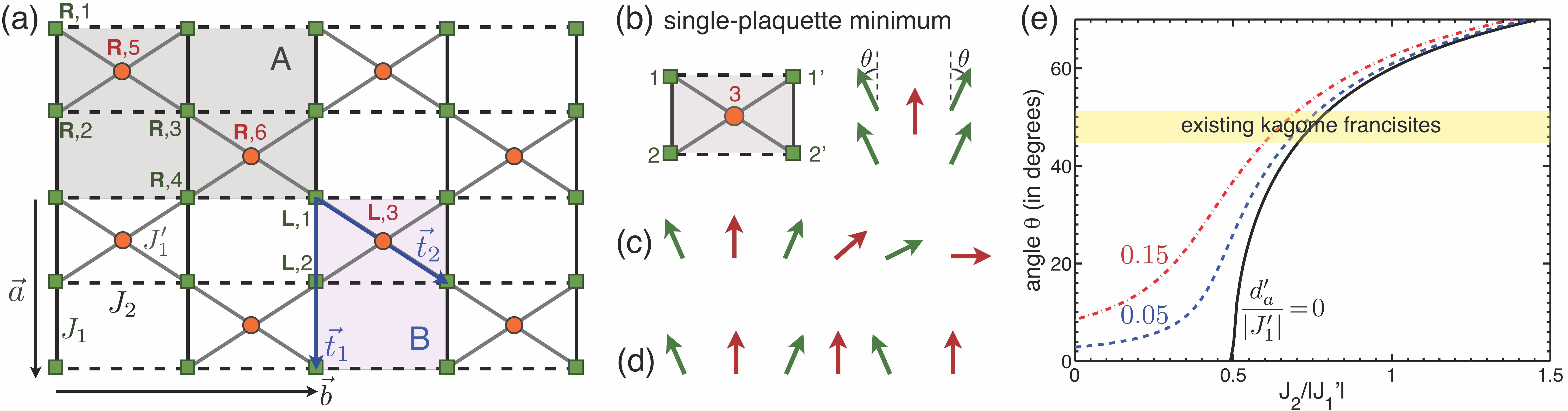}
\caption{(a) Magnetic 2D planes of the kagome francisites, with two symmetry inequivalent Cu sites, Cu1 (green squares) and Cu2 (orange circles). The shaded cells A and B denote the unit cells of the Hamiltonian with and without anisotropy, with primitive translations $\{\vec{a},\vec{b}\}$ and $\{\vec{t}_1\!=\!\vec{a},\vec{t}_2\!=\!(\vec{a}\!+\!\vec{b})/2\}$, respectively. (b) Classical minimum of an isolated crossed plaquette at the isotropic limit. (c-d) Two representative coplanar GSs of the isotropic classical model (there is no modulation along the FM, $J_1$ bonds). (e) The dependence of the in-plane canting angle $\theta$ on $J_2/|J_1'|$ for three different values of $d_a'$, and for all other DM components disregarded. The shaded bar indicates the experimentally observed regime~\cite{pregelj2012}.}
\label{fig:model}
\end{figure*}

\section{Isotropic spin Hamiltonian $\mc{H}_{\sf iso}$}\label{sec:Hiso}
To capture the physics of the full Hamiltonian, it is natural to begin by treating the dominant energy scale first, i.e. the Heisenberg interactions described by $\mc{H}_{\sf iso}$. As it turns out, the isotropic limit is actually very rich since it highlights a number of signatures of highly frustrated magnetism, which is surprising in the light of the dominant FM couplings in the model. Most notably, the isotropic model has a highly degenerate classical GS manifold with a non-trivial continuous structure, that includes both coplanar and non-coplanar states (Sec.~\ref{sec:cgs1}). Moreover, in analogy with the kagome AFM, the degeneracy among the coplanar states survives quantum fluctuations at the harmonic spin-wave level, showing that the eventual order-by-disorder proceeds by non-linear $1/S$ quantum corrections (Sec.~\ref{sec:lswt}). The upshot of this analysis is that the uniform canted phase, which is selected by the non-linear order-by-disorder process, is very fragile even for spins-1/2, as shown explicitly by detailed coupled-cluster (CCM) calculations (Sec.~\ref{sec:GS_select_ccm}).

\subsection{Classical GS manifold}\label{sec:cgs1}
In the absence of magnetic anisotropy, the model has higher translational symmetry than the crystalline symmetry. The unit cell (shaded cell B in Fig.~\ref{fig:model}) comprises three sites and the primitive translations are $\vec{t}_1\!=\!\vec{a}$ and $\vec{t}_2\!=\!(\vec{a}+\vec{b})/2$. The system can be seen as a network of corner-sharing crossed plaquettes, like the one shown in Fig.~\ref{fig:model}\,(b). The classical minimum of this five-site plaquette is the coplanar state 
\bea
\vec{S}_1&=&\vec{S}_2=S\left(\cos\theta\vec{e}_1\!-\!\sin\theta\vec{e}_2\right),~\nonumber\\
\vec{S}_{1'}&=&\vec{S}_{2'}=S\left(\cos\theta\vec{e}_1\!+\!\sin\theta\vec{e}_2\right),~\nonumber\\
\vec{S}_{3}&=&S\vec{e}_1,\nonumber
\eea
where $S=\frac12$ is the spin length, $\vec{e}_1$ is the direction of the total magnetization, $\vec{e}_2\perp\vec{e}_1$, and the canting angle $\theta$ is given by 
\be\label{eq:theta1}
\theta=\left\{
\begin{array}{rr}
\arccos(\frac{-J_1'}{2J_2}), & \text{if} -J_1'\!<\!2J_2\\
0, & \text{otherwise}~.
\end{array}\right.
\ee
A number of remarks are in place here. First, the canting angle depends only on the FM coupling $J_1'$ and not on $J_1$, because $J_2$ does not frustrate $J_1$. Second, the canting does not set in for an infinitesimal $J_2$, i.e. there is a critical value $J_{2}^c=|J_1'|/2$ above which the fully polarized state becomes unstable, see Fig.~\ref{fig:model}\,(e). Finally, the spin plane $\{\vec{e}_1, \vec{e}_2\}$ is arbitrary due to the global SO(3) symmetry of $\mc{H}_{\sf iso}$. In particular, we may rotate the spins $\vec{S}_{3}$ and $\vec{S}_{1',2'}$ around the direction of $\vec{S}_{1,2}$ by any angle $\phi$ at no energy cost.

Having established the GS of a single crossed plaquette, we now proceed to `tile' this solution onto the infinite lattice. To this end, we follow the convention associated with the unit cell B, and label the spin sites by $(\vec{L},\alpha)$, where $\vec{L}\!=\!n\vec{a}\!+\!m\vec{t}_2$, $\alpha\!=\!1$-$3$, and assume open boundary conditions along the $\vec{b}$-axis. It turns out that there is an infinite number of GSs, with two of them depicted in Fig.~\ref{fig:model}\,(c-d) in a 1D representation (as there is no modulation along the $\vec{a}$-axis by virtue of $J_1$). To see the origin of the infinite degeneracy and the structure of the GS manifold, we pick a reference GS (e.g. the canted state of Fig.~\ref{fig:model}\,(d)) and consider the direction $\vec{n}_0\!=\!\vec{S}_{n\vec{a},1-2}$ of the spins along the line $m\!=\!0$. Rotating all spins with $m\!>\!0$ by any angle $\phi_0$ around $\vec{n}_0$ does not change the relative canting angle $\theta$ in any crossed plaquette of the lattice and so it has no energy cost. Similarly, we may rotate all spins with $m\!>\!r$ around the direction $\vec{n}_r\!=\!\vec{S}_{n\vec{a}+r\vec{t}_2,1-2}$ of the spins along the line $m\!=\!r$ by any angle $\phi_r$. Altogether, for any given reference state there is a sub-extensive set of continuous SO(2) rotations defined by $\{\vec{n}_r, \phi_r\}$, each one involving only the spins on the right side of the line $m\!=\!r$. Obviously, any new GS can be used again as a reference state, generating new GSs and so on, leading to an immensely degenerate GS manifold with a non-trivial structure. 

A number of remarks are in order here. First, the GS manifold includes both coplanar and non-coplanar states, and the total magnetization can even vanish for many of the GSs. The coplanar states can be written explicitly as
\be\label{eq:cops}
\vec{S}_{\vec{L}=n\vec{a}+m\vec{t}_2,\alpha}=S\left(\cos\phi_{\alpha,m} \vec{e}_1+\sin\phi_{\alpha,m}\vec{e}_2\right),
\ee
where $\phi_{1,m}\!=\!\phi_{2,m}\!=\!\phi_0\!+\!2\theta\sum_{j=0}^{m-1} q_j$, $\phi_{3,m}\!=\!\phi_{1,m}\!+\!q_m \theta$, $\phi_0$ is arbitrary and $q_j\!=\!\pm 1$. The spiral and the canted states shown in Fig.~\ref{fig:model} (c-d) correspond to the choices $q_j\!=\!1$ and $q_j\!=\!(-1)^j$, respectively. 

Second, the above rotation operations that generate new GSs involve only a fraction of the lattice sites and so they give a new example of so-called `sliding symmetries'~\cite{Nussinov2005,*Nussinov2006}, that are intermediate between global and local. Such gauge-like symmetries appear in several models of strong spin-orbit compounds, where the spin-orbit coupling gives rise to a directional dependence of the effective, Ising-like anisotropic exchange~\cite{KK1973,*KK1982}. Here, these symmetries pertain to the ground state manifold only and the spatial directionality stems from the fact that the $J_2$ couplings run only along the $\vec{b}$-axis, and by the corner-sharing  plaquette structure of the lattice. 

Third, in contrast to the global SO(3) degeneracy, the above degeneracy is accidental, i.e. it is not related to any symmetry of the Hamiltonian. This means that the degeneracy can be lifted by thermal or quantum fluctuations via the 
order-by-disorder effect~\cite{villain80,shender82}, which is discussed in the following.

\begin{figure*}[!t]
\includegraphics[width=0.99\linewidth]{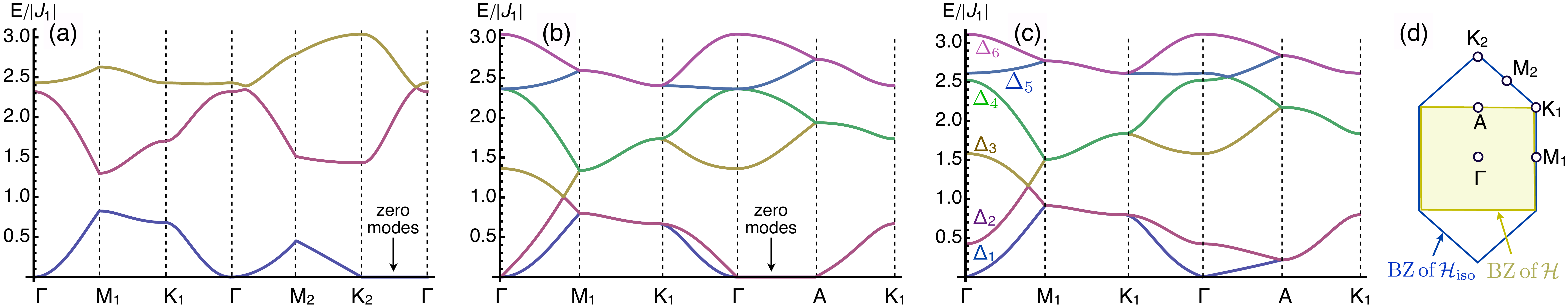}
\caption{(a-b) Linear spin-wave spectra around any coplanar GS of the isotropic model for $J_1\!=\!J_1'$ and $J_2\!=\!0.73|J_1'|$, represented in the BZ (a) of the isotropic model, and (b) of the  crystal (The six spin-wave branches in (b) can be folded back into the three branches of (a)). The spectra in (c) include the effect of a finite DM component $d_a'=0.15|J_1'|$, and will be discussed in Sec.~\ref{sec:LSWSpectraDM}. (d) First BZ's of $\mc{H}_{\text{iso}}$ and $\mc{H}$. 
\label{fig:LSWSpectra}}
\end{figure*}

\subsection{Harmonic fluctuations} \label{sec:lswt}
The existence of an infinite number of coplanar and non-coplanar GSs in the isotropic kagome francisites is strongly reminiscent of what happens in the kagome antiferromagnet~\cite{HKB1992,Chalker1992,Chubukov1992,RCC1993}. There, the non-interacting spin-wave spectrum is the same for all coplanar states, and furthermore there is an entire branch of zero-energy modes, which originate from the presence of an extensive number of soft spin rotations around isolated hexagons~\cite{HKB1992}. As a result, coplanar states are selected at the quadratic level, while non-linear $1/S$ corrections eventually select the so-called $\sqrt{3}\times\sqrt{3}$ ordered state~\cite{Chubukov1992,CCMkagome2011}. (In the extreme quantum limit the so-called $q=0$ state might be favorable, see Ref.~\onlinecite{CCMkagome2011}.)

To find out what happens in the present case we perform a semi-classical expansion around the coplanar GSs. The details of this expansion are provided in App.~\ref{app:lsw1}, and the most relevant results for the present discussion are the following (the remaining aspects of the spectrum will be discussed in Sec.~\ref{sec:LSWSpectraDM} below). First, the non-interacting spin-wave spectrum is again the same for all coplanar states, similar to the kagome antiferromagnet. Second, coplanar states have a line of zero-energy modes along the $\vec{b}$-axis in momentum space, see Fig.~\ref{fig:LSWSpectra}\,(a-b). This structure reveals the existence of a sub-extensive number of soft classical modes that are localized along the $\vec{b}$-axis. It turns out that these modes correspond to tilting three consecutive lines of spins out of the common spin plane: the $\alpha=3$ spins at $m\!=\!r$ by an angle $\xi_3$, the $\alpha\!=\!1$-$2$ spins at $m\!=\!r\!+\!1$ by an angle $\xi_1\!=\!2\xi_3\cos\theta$, and the $\alpha=3$ spins at $m\!=\!r\!+\!1$, again by an angle $\xi_3$. The classical energy cost associated with these modes scales with $\xi_3^4$ instead of $\xi_3^2$, which is why they appear as zero-modes in the quadratic spectrum. 

Despite the fact that there is only a sub-extensive number of soft modes, their presence is strongly suggestive that the coplanar states are most likely selected by quadratic spin-wave corrections. The lifting of the remaining degeneracy among the coplanar states and the eventual selection of the canted phase must thereby proceed via the nonlinear corrections to the theory. In the next section, we shall probe this question by the coupled-cluster expansion method.

\subsection{Beyond harmonic level: Coupled-cluster calculations}\label{sec:GS_select_ccm}
\subsubsection{Key elements of the CCM method}
The coupled-cluster method (CCM) is a universal many-body approach\cite{Bishop98a_CCM} that has been successfully applied to calculate GS properties of frustrated quantum magnets, (see, e.g., Refs.~\onlinecite{ivanov2002magnetic,CCM_LNP2004,darradi2005coupled,schmalfuss2006quantum,zinke2009,farnell2009high,farnell2011honey,CCMkagome2011}). In particular, the CCM can be used to treat non-collinear magnetic ordering such as incommensurate spiral~\cite{darradi2005coupled,zinke2009} as well as commensurate canted phases~\cite{CCMkagome2011,ivanov2002magnetic,farnell2009high}. We will not present details of the CCM methodology here (we refer the interested reader to the references given above), but we shall briefly discuss the key elements of the method. These are the reference (or model) classical state $|\Phi\rangle$ and a complete set of mutually commuting many-body creation operators $\{C_I^+\}$. For convenience, we perform an appropriate rotation of the local axis of the spins such that in the rotated coordinate frame the reference state is a product of spin down states $|\Phi\rangle\!=\!|\!\downarrow\rangle |\!\downarrow\rangle |\!\downarrow\rangle\dots$ The creation operators are then the multispin creation operators $C_I^+\!=\!s_i^+,\,\,s_i^+s_j^+,\,\,s_i^+s_j^+s_k^+,\cdots$, where the indices $i,j,k,\dots$ denote arbitrary lattice sites. The CCM parametrizations of the ket- and bra- GSs read:  
\begin{eqnarray}
\label{ccm}
|\Psi\rangle = e^{\cal S}|\Phi\rangle, 
\qquad 
{\cal S} = \sum_{I \neq 0}{\cal S}_IC_I^+ ; 
\nonumber\\
\langle\tilde{\Psi}| =  \langle\Phi|\tilde{{\cal S}}e^{-{\cal S}},
\qquad 
\tilde{{\cal S}} = 1 + \sum_{I \neq 0}\tilde{\cal S}_IC_I^- ,
\end{eqnarray}
where $C_I^-\!=\!(C_I^+)^+$ and $C_0^+\!\equiv\!1$. Using $\langle\Phi|C_I^+\!=\!0\!=\!C_I^-|\Phi\rangle$, $\forall I\!\neq\!0$, the orthonormality condition $\langle\Phi|C_I^-C_J^+|\Phi\rangle\!=\!\delta_{IJ}$, and the completeness relation 
\begin{equation*}
1=\displaystyle\sum_I C_I^+|\Phi\rangle\langle\Phi|C_I^-=|\Phi\rangle\langle\Phi|+\sum_{I\neq 0}C_I^+|\Phi\rangle\langle\Phi|C_I^-,
\end{equation*}
leads to a set of non-linear and linear equations for the correlation coefficients ${\cal S}_I$ and $\tilde{\cal S}_I$, respectively.

\begin{figure*}[!t]
\includegraphics[width=\linewidth]{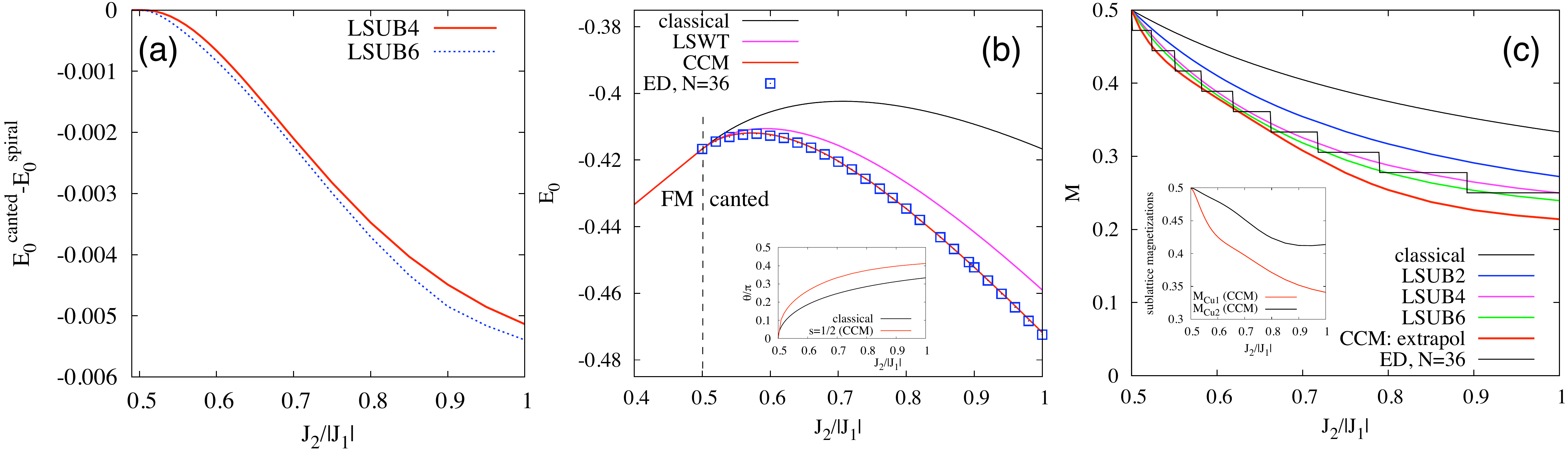}
\caption{\label{fig:CCM}
Key quantities of interest for the isotropic spin-1/2 Heisenberg model at $J_1\!=\!J_1'$, as a function of the frustration parameter $J_2/|J_1'|$.  (a) Difference of the GS energies $\Delta E\!=\!E_0^{\rm canted}\!-\!E_0^{\rm spiral }$ between the canted and the spiral states, as calculated within the CCM LSUB$m$ approximation with $m\!=\!4$ and $6$. (b) GS energy $E_0$ per site calculated by linear spin-wave theory, exact diagonalizations on a periodic $36$-site cluster, and CCM (extrapolated value, see main text). The classical energy is also shown for comparison. Inset: Classical and quantum canting angle $\theta$. (c) Net GS magnetic moment per site $M$ (in units of $g\mu_B\!=\!1$), as calculated by exact digonalizations on the $36$-site cluster, and CCM (LSUB$m$, $m=2,4,6$ and extapolated value, see main text). For comparison we also show the classical value. Inset: Sublattice magnetizations $M_{\text{Cu1}}$ and $M_{\text{Cu2}}$ from CCM (extapolated values). In the classical limit $M_{\text{Cu1}}^{\rm cl}\!=\!M_{\text{Cu2}}^{\rm cl}\!=\!\frac{1}{2}$.}
\end{figure*}

The GS energy is calculated by $E_0\!=\!\langle \Phi|e^{-{\cal S}}He^{\cal S}|\Phi\rangle$, while the expectation values of a physical quantity $A$ (e.g. sublattice magnetizations) is given by $\langle A\rangle\!=\!\langle\tilde\Psi|{\hat A}|\Psi\rangle$, where the Hermitean operator ${\hat A}$ is expressed in the rotated coordinate frame. In the CCM, the only approximation is the truncation of the expansion of the correlation operators ${\cal S}$ and $\tilde{\cal S}$. We use the well established LSUB$m$ scheme, where all multispin correlations on the lattice with $m$ or fewer contiguous sites are taken into account (here, the pairs of sites connected by either $J_1$, $J_1'$, or $J_2$ bonds are all treated as being contiguous sites). The number of these configurations is increasing very rapidly with $m$, and, therefore, we consider here LSUB2, LSUB4, and LSUB6 approximations, only. 

The CCM results can be improved by extrapolating the ``raw'' LSUB$m$ data to $m\!\to\!\infty$. There is ample empirical experience regarding the extrapolation of the GS energy $E_0$ and the sublattice magnetizations $M_\gamma$, $\gamma\!=\!\text{Cu1, Cu2}$. Appropriate extrapolations rules are 
\begin{eqnarray*}
 E_0(m)=a_0+a_1(1/m)^{2}+a_2(1/m)^{4}, \\
 M_\gamma(m)=b_0+b_1(1/m)^{1}+b_2(1/m)^{2}.
\end{eqnarray*} 
Although, the extrapolation improves the results, it is worth mentioning that particularly the results for $M_\gamma$ obtained by the extrapolation with only LSUB2, LSUB4, and LSUB6 data certainly have a limited accuracy.

\subsubsection{Application to isotropic kagome francisites}  
To study the non-linear order-by-disorder process mentioned in Sec.~\ref{sec:lswt}, we apply the CCM method to the two main candidate coplanar reference states: (i) the incommensurate spiral state (Fig.~\ref{fig:model}\,c) with the largest magnetic unit cell, and (ii) the commensurate canted state  (Fig.~\ref{fig:model}\,d) with the smallest magnetic unit cell. Although these states are fundamentally different with respect e.g. to their net total magnetic moment $M$, they have identical GS energies at the harmonic level of spin-wave theory. The CCM expansion can effectively capture the non-linear quantum corrections and thus reveal the GS order-by-disorder process, cf. also the discussion in Ref.~[\onlinecite{CCMkagome2011}]. 

The reference states are characterized by a canting angle $\theta$. As quantum fluctuations may lead to a ``quantum'' canting angle that is different from the classical value, we consider the canting angle in the reference state as a free parameter. We then determine the quantum canting angle by minimizing $E_0^{{\rm LSUB}m}(\theta)$ with respect to $\theta$ for each $m$. For simplicity, we shall take $J_1\!=\!J_1'$, which does not affect the properties of the model qualitatively.

We first discuss the order-by-disorder process.  As shown in Fig.~\ref{fig:CCM}\,(a), the commensurate canted state has lower energy than the spiral state on the quantum $S\!=\!\frac12$ level, and thus non-linear quantum fluctuations select the uniform canted state, which is the state observed experimentally. However, according to Fig.~\ref{fig:CCM}\,(a), the stabilization energy per site appears to be smaller than $0.5$\% of $J_1'$, which shows that the canted phase is very fragile even for spins-1/2. Practically, this means that at finite temperatures the ferrimagnetic order parameter fluctuates very strongly even at short distances, and the system may not attain a long-range magnetic order down to very low temperatures.

Next, we discuss the properties of the canted phase as a function of $J_2$. Figure~\ref{fig:CCM}\,(b) shows the GS energy and the canting angle for the quantum and classical models. The highest energy indicates the strongest competition between $J_2$ and $J_1'$. This maximum is at $J_2/|J_1|\simeq 0.7$ for the classical model and at $J_2/|J_1|\simeq 0.6$ in the quantum case, where the CCM and ED results perfectly match. In the canted phase (i.e. for $J_2\!>\!0.5|J_1|$), the GS energy from LSWT is significantly higher than that obtained by CCM and ED, indicating that for the extreme quantum limit $S\!=\!\frac12$ higher-order than harmonic terms become relevant. The canting angle for the quantum model is always larger than the classical value (note that within the LSWT the classical canting angle is retained).

Figure~\ref{fig:CCM}(c) shows the total ferrimagnetic net magnetic moment $M$ (main panel) as well as the sublattice magnetizations $M_{\rm Cu1}$ and $M_{\rm Cu2}$ (inset) as a function of the frustration parameter $J_2$.  There is a monotonous decrease in both the sublattice and total magnetizations with increasing $J_2$. However, the sublattice magnetizations remain quite large for the parameter range considered here. At $J_2/|J_1|\simeq 0.7$ relevant to the francisites (Table~\ref{tab:exchange}), we expect $M_{\rm Cu1}\simeq 0.86$\,$\mu_B$ and $M_{\rm Cu2}\simeq 0.97$\,$\mu_B$ assuming the powder-averaged $\bar g=2.15$~\cite{millet2001}. These values are in decent agreement with the neutron-scattering results: $M_{\rm Cu1}\simeq 0.92$\,$\mu_B$ and $M_{\rm Cu2}\simeq 0.90$\,$\mu_B$.\cite{pregelj2012} The canting angle $\theta\simeq 60^{\circ}$ is also in reasonable agreement with the experiment ($\theta=51.6^{\circ}$ for \sysBr~\cite{pregelj2012}). However, its value will be further refined in Sec.~\ref{sec:HDM}, where anisotropic terms in the spin Hamiltonian are considered.

\section{Including the DM anisotropy $\mc{H}_{\sf DM}$}\label{sec:HDM}
Having established the magnetism of the isotropic exchange model, we now turn to the impact of the DM anisotropy on the GS structure (Sec.~\ref{sec:cgsdm}), the response under a magnetic field (Sec.~\ref{sec:MvsB}), and the excitation spectrum (Sec.~\ref{sec:LSWSpectraDM}), which are all of direct experimental relevance. Although the DM vectors on the $J_1$ and $J_2$ bonds are very weak in the existent francisites, we shall include them in the discussion in order to draw the most general conclusions. Wherever possible, we provide analytical expressions for a number of experimental quantities, but refrain from explicit comparisons to the experiment until Sec.~\ref{sec:discussion}.

\subsection{Zero-field ground state}\label{sec:cgsdm}
The impact of the DM anisotropy on the zero-field GS can be discussed in two main steps (Secs.~\ref{sec:cgsdm1} and \ref{sec:cgsdm2}), focusing respectively on the in-plane and the out-of-plane canting of the spin structure.

\subsubsection{In-plane canting}\label{sec:cgsdm1}
The DM components along the $\vec{a}$-axis force the spins to lie on the $\vec{b}$-$\vec{c}$ plane but the specific direction of the spins within this plane is still not fixed due to the remaining SO(2) rotation symmetry around the $\vec{a}$-axis. This symmetry will be broken explicitly by the remaining components of the DM vectors.

As we show below, $d_a'$ and $d_{2a}$ stabilize the uniform coplanar canted state of Fig.~\ref{fig:model}\,(d). However, there are two main qualitative differences compared to the isotropic case. First, $d_a'$ and $d_{2a}$  relieve the frustration and select the canted state already at the classical level. Consequently, the stabilization energy is proportional to $d_a'$ and $d_{2a}$, which can be much larger than the corresponding energy gain from the anharmonic order-by-disorder mechanism of the frustrated, isotropic limit, see Fig.~\ref{fig:CCM}\,(a). So, despite the fact that the DM interactions are much weaker than the exchange couplings, the frustration can render the DM anisotropy as the main mechanism for the formation of the canted state. The second difference is that the canting takes place even for infinitesimal $d_a'$ and $d_{2a}$, in contrast to the isotropic limit where the FM state becomes unstable only for $J_2\!\ge\!J_2^c$, see Fig.~\ref{fig:model}\,(e). 

The axes $\vec{e}_1$ and $\vec{e}_2$ of Eq.~(\ref{eq:theta1}) are now in the $\vec{b}$-$\vec{c}$ plane, 
$\vec{e}_1\!=\!\cos\phi\,\vec{b}\!+\!\sin\phi\,\vec{c}$, 
$\vec{e}_2\!=\!-\sin\phi\,\vec{b}\!+\!\cos\phi\,\vec{c}$, 
and the classical canted phase reads
\bea\label{eq:canted}
&&\vec{S}_{1,2}\!=\!S\left[\cos(\phi\!-\!\theta) \vec{b}+\sin(\phi\!-\!\theta) \vec{c}\right], ~~\nonumber\\
&&\vec{S}_{3,4}\!=\!S\left[\cos(\theta\!+\!\phi) \vec{b}+\sin(\theta\!+\!\phi) \vec{c}\right], ~~\\
&&\vec{S}_{5,6}\!=\!S\left(\cos\phi~\vec{b}+\sin\phi~\vec{c}\right)~,\nonumber
\eea
while the total energy $\mc{E}$ is given by
\be\label{eq:Ecanted}
\frac{\mc{E}}{4N_{uc}S^2} \!=\! J_1\!+\!J_2\cos(2\theta)\!+\!2J_1'\cos\theta\!-\!2d_a' \sin\theta\!+\!d_{2a}\sin(2\theta)~,
\ee
where $N_{uc}\!=\!N/6$ is the number of unit cells in the anisotropic model. Note that $\mc{E}$ does not depend on $\phi$ due to the SO(2) symmetry mentioned above. Minimizing with respect to $\theta$ gives
\be\label{eq:theta2}
J_2\sin(2\theta)+J_1'\sin\theta+d_a'\cos\theta-d_{2a}\cos(2\theta)=0,
\ee
which reduces to Eq.~(\ref{eq:theta1}) for $d_a'\!=\!d_{2a}\!=\!0$. Figure~\ref{fig:model}\,(e) 
shows the $J_2$ dependence of the classical canting angle $\theta$ with and without $d_a'$ (for $d_{2a}\!=\!0$). 
As mentioned above, the canting sets in as soon as we switch on an infinitesimal $d_a'$.

\subsubsection{Out-of-plane canting}\label{sec:cgsdm2}
Including the remaining DM components along the $\vec{b}$ and $\vec{c}$ axes, we expect two things to happen. First, the above SO(2) rotation symmetry around the $\vec{a}$-axis is now absent and so the main, FM component of the spins will now point to a specific direction in the $\vec{b}$-$\vec{c}$ plane. Second, the spins will slightly tilt away from the $\vec{b}$-$\vec{c}$ plane. We have investigated the detailed form of this out-of-plane canting by numerical minimizations for the classical model on finite clusters. Disregarding for the moment the much weaker DM components on the $J_1$ and $J_2$ bonds, we find two main qualitative features: (i) the out-of-plane canting involves again only the Cu1 spins and is much smaller than the in-plane canting, and (ii) all five spins in the unit cell point to different directions (unless one of $d_b'$ or $d_c'$ equals to zero, see below), i.e. the magnetic structure of a single kagome layer has five sublattices, compared to three in the $d'_{b,c}\!=\!0$ case. 

The simplest way to understand the above out-of-plane canting structure is to take the coplanar state of the $d'_{b,c}\!=\!0$ limit and evaluate the contributions to the local fields $\vec{B}_\alpha(d_{b,c}')$ exerted by $d_b'$ and $d_c'$. We find:
\bea\label{eq:localBs}
\vec{B}_{1}(d_{b,c}') \!&=&\! -\vec{B}_{2}(d_{b,c}') \!=\! 2S\left[-d_b'\sin\phi \!+\! d_c'\cos\phi \right]~\vec{a},~~~\nonumber\\
\vec{B}_{3}(d_{b,c}') \!&=&\! -\vec{B}_{4}(d_{b,c}') \!=\! 2S\left[ d_b'\sin\phi \!+\! d_c'\cos\phi \right]~\vec{a},~~~\\
\vec{B}_{5}(d_{b,c}') \!&=&\!\vec{B}_{6}(d_{b,c}') \!=\! 0~. \nonumber
\eea
These expressions deliver a number of clear insights: (i) the Cu2 spins (5 and 6) remain on the $\vec{b}$-$\vec{c}$ plane; (ii) spins 1 and 2 tilt away from the $\vec{b}$-$\vec{c}$ plane but in opposite directions; (iii) spins 3 and 4 also tilt in opposite directions; (iv) the relative amount of tilting between 1-2 and 3-4 as well as the global angle $\phi$ of the total moment within the $\vec{b}$-$\vec{c}$ plane are fixed by the competition between $d_b'$ and $d_c'$. The resulting five-sublattice magnetic structure can be described by 
\bea\label{eq:oopansatz}
&&\vec{S}_{1,2}/S\!=\!\cos\gamma_{1}\left[ \cos(\phi\!-\!\theta) \vec{b}\!+\!\sin(\phi\!-\!\theta) \vec{c} \right] \!\pm\! \sin\gamma_{1}\vec{a}, ~~\nonumber\\
&&\vec{S}_{3,4}/S\!=\!\cos\gamma_{3}\left[ \cos(\theta\!+\!\phi) \vec{b}\!+\!\sin(\theta\!+\!\phi) \vec{c} \right] \!\pm\! \sin\gamma_{3}\vec{a}, ~~~~~~~\\
&&\vec{S}_{5,6}/S\!=\!\cos\phi~\vec{b}\!+\!\sin\phi~\vec{c}~,\nonumber
\eea
where the out-of-plane canting angles $\gamma_1$ and $\gamma_3$, together with $\theta$ and $\phi$, can be determined by minimizing the total energy, see App.~\ref{app:oopansatz}. For the special case of $d_c'\!=\!0$ (respectively $d_b'\!=\!0$) we find $\gamma_1\!=\!-\gamma_3$ (resp. $\gamma_1\!=\!\gamma_3$), and $\phi\!=\!90^\circ$  (resp. $\phi\!=\!0^\circ$), i.e., the total moment points along the $\vec{c}$ (resp. $\vec{b}$) axis.

The existent francisites have $\phi\!=\!\pi/2$, which means that $d_{c}'\!\ll\!d_{b}'$  (see below). In this case, we may also show that the much weaker DM components on the $J_1$ and $J_2$ bonds favor the same out-of-plane canting structure, since their contributions to the local fields also obey (\ref{eq:localBs}). The out-of-plane canting angles can be found analytically (see App.~\ref{app:oopansatz}):
\be\label{eq:gamma13}
\gamma_1=-\gamma_3=\frac{d_b'-d_{1b}\cos\theta}{2J_1+J_1' \sec\theta+d_{2a}\tan\theta}~.
\ee
Note that, unlike the in-plane canting case, there is now an explicit dependence on the FM coupling $J_1$, because the out-of-plane canting comes with a finite exchange energy cost on each $J_1$ bond. 

Finally, it is worth summarizing the symmetry properties of the above zero-field GS. Clearly, this state does not break the translational symmetry in the $ab$-plane, but there is a doubling of the unit cell along the $\vec{c}$-axis on account of the AFM interlayer coupling $J_{\perp,2}$ (see below). Furthermore, the GS breaks the two orthogonal reflection planes going through the Cu2 sites, but preserves the inversion symmetry around the Cu1 sites, as well as the glide plane operations that involve the  non-primitive translations along the two diagonals of the lattice, $(\vec{a}\pm\vec{b})/2$, followed by a reflection in the $ab$-plane. As we discuss in Sec.~\ref{sec:LSWSpectraDM} below, the latter symmetry gives rise to a two-fold degeneracy for all spin-wave modes along two special lines in momentum space.


\begin{figure*}[!t]
\includegraphics[width=1\linewidth]{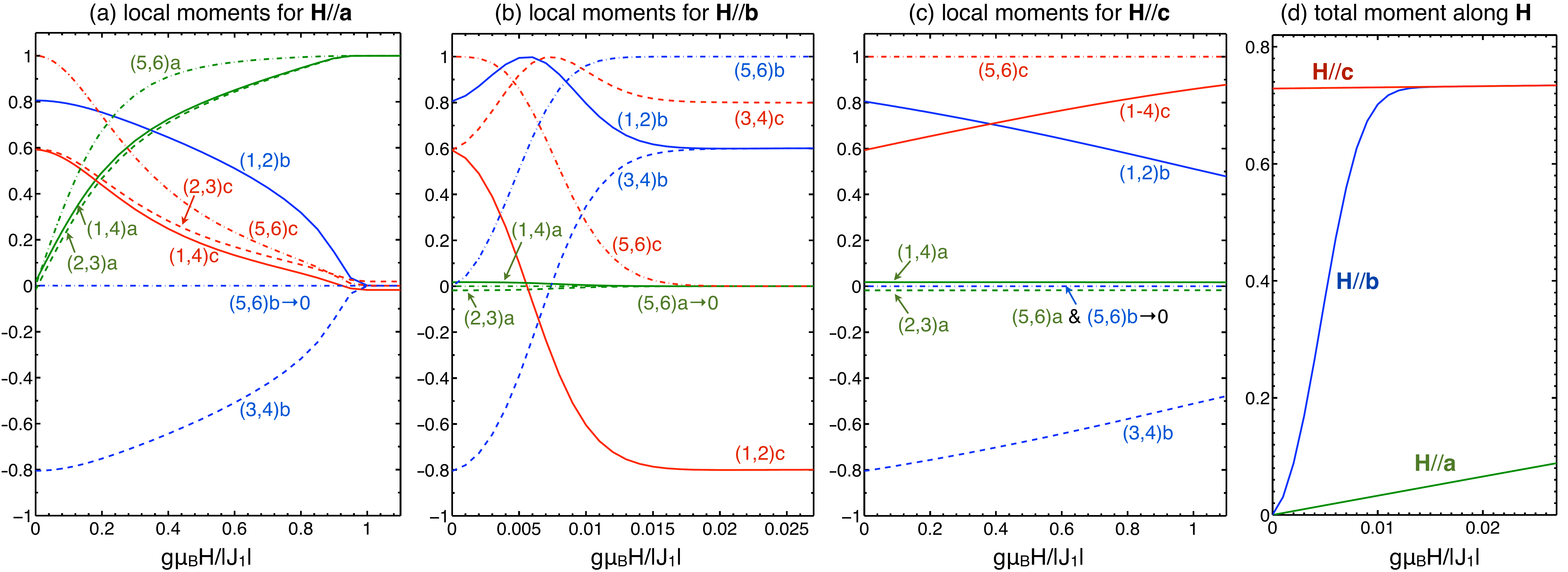}
\caption{\label{fig:MvsH} Anisotropic response of the kagome francisites under a magnetic field $\vec{H}$. Local moments (a-c) and total magnetization per site (d)  for three different directions of $\vec{H}$ (in units of $\mu_B$, for $g\!=\!2$). The results are obtained by numerical minimizations of the total energy of a finite 2D layer with 8x8x6 sites and periodic boundary conditions. The microscopic parameters used are (in K): $J_1=-75$, $J_1'=-67$, $J_2=49$, $d_a'=12.1$, $d_b'=-4.7$, $d_c'=0$, and we have disregarded the much weaker DM terms from the $J_1$ and $J_2$ bonds. Notation: `(1,4)a' stands for the local moments of sites 1 and 4 of Fig.~\ref{fig:structure} along the $\vec{a}$ axis, etc.}
\end{figure*}

\subsection{Ground state evolution in a magnetic field}\label{sec:MvsB}
The DM anisotropy has direct consequences in the response of the system under a magnetic field $\vec{H}$. The qualitative aspects of this response are shown in Fig.~\ref{fig:MvsH}, which shows data from finite-size numerical minimizations of our spin model in the classical limit. The results are all in agreement with the anisotropic response reported for the bromide and chloride compounds~\cite{pregelj2012,miller2012}. Let us discuss the main features in some detail.

\subsubsection{Magnetization process for $\vec{H}\!\parallel\!\vec{a}$}\label{sec:MvsBa}
A magnetic field along the $\vec{a}$-axis turns the spins out of the $\vec{b}$-$\vec{c}$ plane, and thus acts against the DM components $d_a'$ and $d_{2a}$. In contrast to the $\vec{H}\!\parallel\!\vec{c}$ case (see below), there is no metamagnetic transition since the rotation of the spins out of the $\vec{b}$-$\vec{c}$ plane can take place continuously in each layer. We can describe the field-induced out-of-plane canting by a slight modification of  Eq.~(\ref{eq:canted}) for $\phi\!=\!\pi/2$:
\bea\label{eq:B//a}
&&\vec{S}_{1,2}\!=\!S\left[\cos\beta_1\left(\sin\theta_H\vec{b}+\cos\theta_H\vec{c}\right)+\sin\beta_1 \vec{a}\right], ~~\nonumber\\
&&\vec{S}_{3,4}\!=\!S\left[\cos\beta_1\left(-\sin\theta_H\vec{b}+\cos\theta_H\vec{c}\right)+\sin\beta_1\vec{a}\right], ~~~~\\
&&\vec{S}_{5,6}\!=\!S\left[\cos\beta_5\vec{c}+\sin\beta_5\vec{a}\right]~,\nonumber
\eea
where $\beta_{1,5}$ are the field induced out-of-plane canting angles, and we have used the symbol $\theta_H$ for the field dependent in-plane canting angle between neighboring Cu1 and Cu2 sites. Here we are assuming that the field is such that the out-of-plane canting driven by $d_b'$ and $d_{1b}$ alone is infinitesimal compared to that driven by the field itself, and so we disregard $d_b'$ and $d_{1b}$. This condition is satisfied already for very small fields, see Fig.~\ref{fig:MvsH}(a).

The angles $\beta_{1,5}$ and $\theta_H$ can be found analytically at small fields by minimizing the total energy. To leading order in the field, the angle $\theta_H$ remains unaffected by the field, i.e., it is given by (\ref{eq:theta2}), while $\beta_{1}$ and $\beta_5$ grow linearly with the field, and the same holds for the total magnetization along the field, see App.~\ref{app:slopea} and Fig.~\ref{fig:MvsH} (d). The corresponding slope $\kappa_a$ of the magnetization per site at small fields is given by
\be\label{eq:slope-a}
\kappa_a\!=\!\frac{(g\mu_B)^2}{12}\frac{J_1'(1+2\cos\theta)^2\csc\theta\!+\!d_{2a}-4d_a'\cos\theta}{d_a'(J_1'+d_{2a}\sin\theta)-J_1' d_{2a} \cos\theta}.
\ee
Note that $\kappa_a$ diverges for $d_a', d_{2a}\to 0$, which is physically expected since, in the absence of the DM anisotropy, an infinitesimal field along $\vec{H}\!\parallel\!\vec{a}$ will immediately rotate the total moment along $\vec{a}$.

In the bromide compound~\cite{pregelj2012}, the magnetization follows an almost linear behavior up to 8~T, with a much smaller slope than the one shown from the low-field response along $\vec{b}$. This is a clear confirmation that $d_a'$ is much larger than $d_b'$, in agreement with our {\it ab initio} results (Table~\ref{tab:dm}).

\subsubsection{Magnetization process for $\vec{H}\!\parallel\!\vec{b}$}\label{sec:MvsBb}
A magnetic field along the $\vec{b}$-axis acts against the DM components $d_b'$ and $d_{1b}$, since it globally rotates the total moments from $\vec{c}$ to $\vec{b}$ and, at the same time, diminishes the out-of-plane canting angles $\gamma_1$ and $\gamma_3$, see Fig.~\ref{fig:MvsH} (b). As for the case of $\vec{H}\!\parallel\!\vec{a}$, there is again no metamagnetic transition since the rotation of the moments from $\vec{c}$ to $\vec{b}$ can take place continuously, with opposite sense of rotation in subsequent layers. The rotation will be completed at a characteristic field strength $H_{b,1}$ which is set by the weak DM components $d_b'$ and $d_{1b}$. Above $H_{b,1}$, $\phi\!=\!0$, $\gamma_{1,3}\!=\!0$, and the behavior of the total moment becomes identical to the corresponding response along the $\vec{c}$ axis, see Fig.~\ref{fig:MvsH} (d).

The value of $H_{b,1}$ contains information for the strength of the DM components $d_b'$ and $d_{1b}$. To see this we note that the GS configuration adjusts adiabatically under an infinitesimal change of the field, and the internal energy change is given by the external work done by the field, $d\mc{E}\!=\!\sum_i \frac{\partial\mc{E}}{\partial\vec{S}_i}\!\cdot\!d\vec{S}_i\!=\!\vec{H}\!\cdot\!d\vec{M}$, where we have used the minimum energy conditions $\frac{\partial}{\partial\vec{S}_j}\left(\mc{E}\!-\!\vec{M}\!\cdot\!\vec{H}\right)\!=\!0$, and $\vec{M}\!=\!-g\mu_B\sum_i\vec{S}_i$ is the total magnetization. The total internal energy change from the state at $H\!=\!0$ to the state at $H\!=\!H_{b,1}$ is then given by
\be\label{eq:W}
\Delta \mc{E}_b = \mc{E}(H_{b,1})-\mc{E}(0)= \int \vec{H}\cdot d\vec{M}\equiv W_b~.
\ee
Experimentally, $H_{b,1}\!\simeq\!7$~T for the bromide compound~\cite{pregelj2012}, and the external work can be found by integrating the measured $H$ vs.\! $M$ curves, see e.g. Fig.~2~(b, inset) of Ref.~[\onlinecite{pregelj2012}]. On the other hand, the internal energy change can be found from our spin model, by making the approximation (based on $d_b', d_{1b}\!\ll\!J_2$) that the field does not affect the canting angle $\theta$ up to $H_{b,1}$. We find $\Delta \mc{E}_b /N \!=\! \frac{4}{3} S^2 \left(-d_b'\!+\!d_{1b}\cos\theta\right)\gamma_1(H\!=\!0)$, which, in conjunction with (\ref{eq:W}) and (\ref{eq:gamma13}) gives
\be
\frac{W_b}{N}=\frac{4S^2}{3}\frac{(-d_b'+d_{1b}\cos\theta)^2}{-2J_1+2J_2+d_a' \csc\theta-d_{2a} \cot\theta}~.
\ee
As expected, the external work vanishes for $d_b'=d_{1b}=0$.

\subsubsection{Magnetization process for $\vec{H}\!\parallel\!\vec{c}$}\label{sec:MvsBc}
Here, a magnetic field $\vec{H}\!\parallel\!\vec{c}$ acts against the AFM coupling $J_2$ (i.e., it reduces the canting angle $\theta$) but also against the interlayer AFM coupling $J_{\perp2}$. Because of the latter, there is a first-order, metamagnetic transition at a characteristic field $H_{c,1}$, which corresponds to the Zeeman energy required to turn the total moments in every second $ab$-layer against $J_{\perp2}$. At $H_{c,1}$, the total moments point along $+\vec{c}$ in all layers, i.e., we enter a ferrimagnetic configuration with five sublattices in total. Experimentally, $H_{c,1}\!\simeq\!0.8$~T for both X=Br and Cl compounds~\cite{pregelj2012,miller2012}.

Figure~\ref{fig:MvsH} (c) shows the calculated response for a single layer, which should describe the behavior of the system above $H_{c,1}$. The response is described by the ansatz of Eq.~(\ref{eq:oopansatz}), with $\phi\!=\!\pi/2$, but now the angle $\theta$ depends on the field strength. Eventually this angle will vanish when we reach the fully polarized state, at a characteristic field which is set by $J_2$. A straightforward minimization of the classical energy shows that the magnetization increases linearly with the field above $H_{c,1}$, see Fig.~\ref{fig:MvsH} (c, d) and App.~\ref{app:slopec}. The corresponding slope $\kappa_c$ of the magnetization per site is given by
\be\label{eq:slopec}
\kappa_c=\frac{(g\mu_B)^2/3}{2J_2+[d_a'-d_{2a} \cos\theta (1+2\sin^2\theta)]/\sin^3\theta},
\ee  
where all couplings are in units of energy. Note that in the limit $d_a', d_{2a}\to 0$, the slope becomes inversely proportional to the AFM coupling $J_2$, which is the expected behavior.
\\


\subsection{Excitation spectrum}\label{sec:LSWSpectraDM}
Besides the qualitative impact on the zero-field GS and the anisotropic response in a magnetic field, the DM anisotropy has also qualitative consequences in the magnon excitation spectrum. To demonstrate this aspect we perform a linear spin-wave expansion around the canted state of Eq.~(\ref{eq:canted}) with $\phi\!=\!\pi/2$. The details of this calculation are provided in App.~\ref{app:lsw2}. Here, we shall discuss the results in two separate steps, focusing respectively on the influence of the DM components along the $\vec{a}$-axis and those along the $\vec{b}$-axis. The DM components along the $\vec{c}$-axis can be disregarded as discussed above. For comparison to experiments, the formulas given below include the Zeeman contribution from an applied magnetic field along the $\vec{c}$-axis. Finally, we shall also comment on the influence of the interlayer coupling $J_{\perp2}$.

\subsubsection{Influence of $d_a'$ and $d_{2a}$}
Figure~\ref{fig:LSWSpectra}\,(c) shows the dispersions of the six spin-wave branches of the system along some representative symmetry directions of the Brillouin zone (BZ) for $J_1\!=\!J_1'$, $J_2\!=\!0.73|J_1'|$, and $d_a'\!=\!0.15|J_1'|$, with all other DM components neglected. Comparing to the isotropic case of Fig.~\ref{fig:LSWSpectra}\,(b), there are two main effects driven by $d_a'$. First, the line of zero modes along the $\vec{b}$-axis in momentum space is gapped out, which is a direct consequence of the fact that the DM component $d_a'$ lifts the non-trivial degeneracy of the GS manifold and selects the coplanar canted state already at the classical level. The second effect of $d_a'$ is that one of the two Goldstone modes at the $\bs{\Gamma}$ point of Fig.~\ref{fig:LSWSpectra}\,(b) acquires a finite spin gap $\Delta_{\Gamma,2}$. The other mode, $\Delta_{\Gamma,1}$, remains gapless on account of the global SO(2) continuous symmetry around the $\vec{a}$-axis. This mode will be eventually also gapped out by the remaining DM components (see below) but also by the demagnetizing fields~\cite{Kittel1947,*Kittel1948} which become relevant e.g. for $H\!>\!H_{c,2}$ when $\vec{H}\!\parallel\!\vec{c}$.

In the absence of $d_b'$ and $d_{1b}$, the excitation structure at the $\bs{\Gamma}$ point can be worked out analytically for the four modes $\Delta_{\Gamma,2}$, $\Delta_{\Gamma,3}$, $\Delta_{\Gamma,4}$ and $\Delta_{\Gamma,6}$, see App.~\ref{app:lsw2} and Fig.~\ref{fig:LSWSpectra}\,(c). The mode at $\Delta_{\Gamma,3}$ corresponds to an out-of-phase rotation of the two Cu2 spins of the unit cell, with
\be\label{eq:Delta3}
\Delta_{\Gamma,3}=4S(-J_1'\cos\theta_H+d_a' \sin\theta_H)+g_c\mu_B H~~.
\ee
This mode does not involve the Cu1 spins, which explains why $\Delta_{\Gamma,3}$ does no depend on the Cu1-Cu1 couplings $J_1$, $J_2$, and $d_{2a}$. The modes at $\Delta_{\Gamma,2}$, $\Delta_{\Gamma,4}$ and $\Delta_{\Gamma,6}$, on the other hand, involve only the Cu1 spins, and are given by the exact expressions:
\begin{widetext}
\bea\label{eq:Delta246}
&&\Delta_{\Gamma,2}^2=(d_a'\csc\theta_H\!-\!d_{2a}\cot\theta_H\!+\!h_c \cos\theta_H) [2J_2\sin^2\theta_H\!+\!d_a'\csc\theta_H\!-\!d_{2a}(\cot\theta_H\!+\!\sin(2\theta_H))\!+\!h_c \cos\theta_H],\\
&&\Delta_{\Gamma,4}^2=(-2J_1\!+\!d_a'\csc\theta_H\!-\!d_{2a}\cot\theta_H\!+\!h_c \cos\theta_H) [2J_2\sin^2\theta_H\!-\!2J_1\!+\!d_a'\csc\theta_H\!-\!d_{2a}(\cot\theta_H\!+\!\sin(2\theta_H))\!+\!h_c \cos\theta_H],~~~~~~~\\
&&\Delta_{\Gamma,6}^2=[2J_1\!-\!2J_2\!+\!J_2\sin^2\theta_H\!-\!d_a'\csc\theta_H \!+\! d_{2a}(\cot\theta_H\!-\!\sin(2\theta)/2)\!+\!h_c \cos\theta_H]^2 \!-\! \sin^2\theta_H (J_2\sin\theta_H \!-\! d_{2a}\cos\theta_H)^2~,
\eea
\end{widetext}
where we have defined $h_c\equiv g_c\mu_BH$.
We note that $\Delta_{\Gamma,2}$ does not depend on $J_1$,  because in the corresponding mode spins that are coupled by $J_1$ rotate in phase with each other, see App.~\ref{app:lsw2}. It is worth discussing the leading behavior of the above expressions at zero field and for $d_a'$, $d_{2a}\ll |J_1|$, $J_2$:
\bea\label{eq:Delta246b}
&&\Delta_{\Gamma,2} \simeq \sqrt{2J_2 \sin\theta (d_a'\!-\!d_{2a}\cos\theta)}, \\
&&\Delta_{\Gamma,4} \simeq 2\sqrt{-J_1 (-J_1+J_2\sin^2\theta)}, \\
&&\Delta_{\Gamma,6}\simeq 2\sqrt{(-J_1+J_2)(-J_1+J_2\cos^2\theta)}~.
\eea
The most striking feature of the above expression for $\Delta_{\Gamma,2}$ is that it does not scale with $d_a'$ but with the geometric mean of the AFM coupling $J_2$ and the DM anisotropy, which is reminiscent of the corresponding formula for the spin-wave gap in uniaxial antiferromagnets~\cite{Nagamiya1951,Kittel1951,KefferKittel1952}. As a result, the gap opening can be much larger than anticipated based on the DM energy scale alone.

\subsubsection{Apparent $g$-factors in a magnetic field $\vec{H}\parallel\vec{c}$}
As shown in App.~\ref{app:Gmodes}, the mode at $\Delta_{\Gamma,3}$ involves only the Cu2 sites and so the Zeeman energy contribution to this mode is $g_c\mu_B H$. By contrast, the modes at $\Delta_{\Gamma,2}$, $\Delta_{\Gamma,4}$ and  $\Delta_{\Gamma,6}$ involve only the Cu1 spins, which are canted by an angle $\theta_H$ from the direction of the field, and so the Zeeman contribution to the energy of these modes is $g\mu_B \cos\theta_H H$.However, one should keep in mind that the angle $\theta_H$ changes with the field too, so the behavior under a magnetic field is not given by the simple Zeeman expressions, but is more involved (see above expressions). For example, for the mode at $\Delta_{\Gamma,3}$, the apparent slope is 
\be
\frac{g_{\Gamma,3}}{g_c} \!=\! 1\!+\!2\cos\theta\!+\!\frac{d_{2a}\!-\!d_a'\cos\theta}{J_2\sin^3\theta\!+\!\frac{1}{2}d_a'\!+\!\frac{1}{4}d_{2a}[\cos(3\theta)\!-\!3\cos\theta]},
\ee
as can be found using (\ref{eq:Delta3}) and Eq.~(\ref{eq:epsilon}) for the leading field dependence of $\theta_H$.

\subsubsection{Influence of $d_b'$ and $d_{1b}$}
To gauge how strong is the dependence of $\Delta_{\Gamma,1}$ on the DM components along the $\vec{b}$-axis, we perform a direct numerical calculation of the spin-wave spectrum in the presence of finite $d_b'$ and $d_{1b}$ (unfortunately, analytical results are not possible for this case). Using typical numbers for the exchange and DM couplings gives $\Delta_{\Gamma,1}\!<\!1$~K, which has the energy scale of the demagnetizing field contributions to the gap. So, unlike $\Delta_{\Gamma,2}$ which grows quickly with $d_a'$ and $d_{2a}$, the first spin gap $\Delta_{\Gamma,1}$ grows much more slowly with $d_b'$ and $d_{1b}$. This is consistent with the ESR data reported for the bromide compound in Ref.~[\onlinecite{millet2001}], where the line assigned to the bulk (line `B') shows no observable spin-wave gap.

Finally, we should emphasize one aspect of the spectrum which remains robust even in the presence of all DM components. Namely, that the six spin-wave modes pair up into three spin-wave branches along the special directions $\vec{k}\!=\!(k_a,\pi)$ (line A-K$_1$ in Fig.~\ref{fig:LSWSpectra}\,(d)) and $\vec{k}\!=\!(\pi,k_b)$ (line M$_1$-K$_1$). As we show explicitly in App.~\ref{app:glides}, this degeneracy stems from the glide plane symmetries that involve non-primitive translations along $(\vec{a}\pm\vec{b})/2$, followed by a reflection in the $ab$-plane. As discussed above, these symmetries remain unbroken even in the presence of all DM couplings. A uniform magnetic field in the $ab$-plane will break this symmetry. The same is actually true for a field along the $\vec{c}$-axis due to the finite off-diagonal element $a_{23}$ of the $\vec{g}$-tensor of the Cu1 sites, see Sec.~\ref{sec:anis}.

\subsubsection{Influence of $J_{\perp2}$}
As discussed above, at zero-field the direction of the total moment alternates between $\vec{c}$ and $-\vec{c}$ from one $ab$-layer to the next, i.e., there is a doubling of the magnetic unit cell along the crystallographic $\vec{c}$-axis. This gives rise to an additional weak dispersion of the six spin-wave branches along the $\vec{c}$ axis in momentum space. This dispersion eventually disappears for fields $\vec{H}\!\parallel\!\vec{c}$ above the metamagnetic transition, where the total moments of all kagome layers are aligned with each other.

\section{Discussion and Summary}
\label{sec:discussion}
The canted order of the kagome francisites corresponds to an unconventional ferrimagnetism. Conventional ferrimagnets are built by antiferromagnetically interacting nonequivalent sublattices. The ferrimagnetic GS is a collinear state, and the magnitude of the net magnetization is a simple fraction of the saturated magnetization. This kind of ferrimagnetism is often called Lieb-Mattis ferrimagnetism~\cite{lieb62, ferri_Ivanov, ferri_Nakano, Jagannathan2006, ioannis1,*ioannis2}. Francisites exhibit a frustration-induced unconventional (non-Lieb-Mattis) ferrimagnetism, where the GS is noncollinear and the net magnetization is not a simple fraction of the saturated magnetization~\cite{ferri_Nakano,ivanov2002magnetic}.

The first main result of the present study is that this magnetism cannot be captured by isotropic Heisenberg interactions alone in the case of the  kagome francisites. While the local canting of the moments can be ascribed to the frustrating AFM coupling $J_2$, the question about long-range ordering is much richer because of the kagome topology of the spin lattice. Despite the fact that these systems are very close to the fully polarized state, the lattice topology gives rise to an infinite degeneracy of the classical ground state manifold with a non-trivial structure that includes both coplanar and non-coplanar states. Surprisingly, this degeneracy is only lifted by non-linear $1/S$ spin-wave corrections and, although the selected phase is most likely the uniform canted phase observed experimentally, the stabilization energy associated with the non-linear order-by-disorder process is only about 0.5\% of the dominant FM couplings, as shown by extensive CCM calculations. As a result these systems would fluctuate over an infinite manifold of coplanar states and would not order down to very low temperatures, which is at odds with the relatively high $T_N\!\simeq\!25$~K. Including the microscopic DM interactions resolves this puzzle and moreover explains all qualitative properties of the kagome francisites, most notably the detailed structure of the ground state and the anisotropic response in a magnetic field. 

Let us now juxtapose the main theoretical picture with the experimental results reported so far. Experimentally, spins lie in the $bc$ plane with the net moment along the $\vec{c}$ direction. DFT results suggest that this configuration is caused by $d_a'$, the leading $a$-component of the DM vector on the $J_1'$ bonds, but this component leaves the freedom of rotating spins in the $bc$ plane and does not fix the direction of the net moment. The alignment of the net moment with the $c$-axis is driven by weaker DM components and is related to the out-of-plane canting. This canting is qualitatively consistent with the experiment, because Pregelj~\textit{et al.}~\cite{pregelj2012} use the $\Gamma_3$ irreducible representation, which indeed allows the canting of this type. However, they do not find any $a$-component of the magnetic moment in zero field, whereas at 1\,T for $H\|\vec{c}$, above the metamagnetic transition, the $a$-component is determined with a large error bar, $-0.2(1)$\,$\mu_B$.

In order to stabilize the out-of-plane canting and direct the net moment along $\vec{c}$, we have to assume that $d_c'\ll d_b'$, which is at odds with the DFT result $d_c'\simeq d_b'$ (Table~\ref{tab:exchange}). Taking $d_c'\ll d_b'$ as an \textit{ad hoc} assumption, we derive the magnetization process for different field directions (Sec.~\ref{sec:MvsB}) and achieve an  excellent qualitative agreement with the experiment. Magnetic field applied along $\vec{c}$ triggers a metamagnetic transition, and a large net moment appears abruptly. When the field is applied along other directions, no metamagnetic transition is observed. The system is gradually polarized, and the slope for $H\|\vec{b}$ is much higher than for $H\|\vec{a}$, see Fig.~\ref{fig:MvsH}(d). This perfectly matches experimental results\cite{pregelj2012,miller2012} that demonstrate the abrupt increase in the magnetization for $H\|\vec{c}$, a gradual (and slower) increase for $H\|\vec{b}$ and an even slower growth of the magnetization for $H\|\vec{a}$. This pronounced magnetic anisotropy is well explained by the proposed arrangement of the DM vectors.

\begin{table}
\caption{\label{tab:comparison}
Experimental parameters for \sysBr\ (Ref.~\onlinecite{pregelj2012}) and their expected values based on DFT estimates of individual exchange couplings (Table~\ref{tab:exchange}). Listed are: {the Curie-Weiss temperature $\Theta_a$ for $\vec{H}\!\parallel\!\vec{a}$ (see App.~\ref{app:MF}), the} in-plane canting angle in the $bc$ plane $\theta$, the out-of-plane canting angle for Cu1 spins $\gamma_1$, the slope of the magnetization curves $\kappa_a$ and $\kappa_c$ for $H\|\vec{a}$ and $H\|\vec{c}$, respectively, and the work (per site) $W_b/N$ done by the field when $H\|\vec{b}$ from $H=0$ up to $H_{b,1}$ (see text). The latter corresponds to the parameters of Fig.~\ref{fig:MvsH}.}
\begin{ruledtabular}
\begin{tabular}{ccccccc}
           &$\Theta_a$&  $\theta$ & $\gamma_1$ & $\kappa_a$ & $\kappa_c$ & $W_b$ \\
           &(K) &    (deg)  &   (deg)    & ($\mu_B$/T)& ($\mu_B$/T)&  ($\mu_B$T)   \\\hline
Experiment &60&     51.6  &    0       &    0.066   &   0.0074   &  2.9   \\
Theory     & 53&    52.1  &    0.7     &    0.16    &   0.0060   &     0.2\\
\end{tabular}
\end{ruledtabular}
\end{table}

It is now tempting to make a quantitative comparison between our DFT-based microscopic magnetic model and the anisotropic magnetic response probed experimentally. In Table~\ref{tab:comparison}, we summarize key experimental quantities and their predictions based on the DFT estimates of individual couplings. We find very good agreement for the Curie-Weiss temperature $\Theta_a$ for $\vec{H}\!\parallel\!\vec{a}$ (this is the direction along which the Curie-Weiss temperature depends very little on the DM anisotropy, see App.~\ref{app:MF}), the canting angle $\theta$, reasonable values for $\kappa_c$ and $\gamma_1$ (experimentally, $\gamma_1\!=\!0$ in zero field and $\gamma_1\!=\!11(6)^{\circ}$ at 1\,T), and a very poor match for $\kappa_a$ and the work $W_b$ done by the field $\vec{H}\!\parallel\!\vec{b}$ up to $H\!=\!H_{b,1}$. This may reflect the fact that $\theta$ is largely dependent on the isotropic couplings $J_i$, which are determined from DFT with the much higher accuracy than the DM couplings that control the magnetic anisotropy. It would be natural to make a reverse procedure and determine individual microscopic parameters from the experimental quantities, but, unfortunately, the problem is largely under-determined because of the large number of individual DM parameters. Moreover, the experimental value of $\gamma_1$ is determined with a very low precision.

The analysis of magnetic excitations reveals even more acute discrepancies between our model and the experiment. Wang~\textit{et al.}~\cite{wang2012} report two magnetic modes around 14\,K (1.25\,meV) in X=Br. Miller~\textit{et al.}~\cite{miller2012} observed a presumably similar mode at 13.7\,K (9.5\,cm$^{-1}$) and another one around 47\,K (33.1\,cm$^{-1}$) in X=Cl. Therefore, the low-lying mode(s) around 14\,K seem to be generic for francisites, but they fail to find any explanation in our model, because the energy of the lowest mode $\Delta_{\Gamma,1}$ remains below 1\,K for the parameter regime considered, while $\Delta_{\Gamma,2}$ lies at a higher energy of about 29\,K and might match the experimental mode at 47\,K. Note also that the modes at 14 and 47\,K clearly have a different origin because they show different evolution in the magnetic field. The former mode yields $g_{\rm eff}\simeq 2.16$ that would be typical for Cu2, whereas the five times lower slope~\cite{miller2012} of the 47\,K mode indicates the predominant contribution of Cu1.

Altogether, we are able to achieve a qualitative microscopic description of francisites, but several quantitative features point to a more complex nature. The deficiencies of our model may be partly related to inaccuracies of DFT, especially when weak DM couplings are considered, but we believe that experimental uncertainties are substantial as well. Several experimental studies allude to a structural phase transition in both Cl and Br compounds at $100$-$150$\,K,\cite{millet2001,miller2012} although the detailed nature of these transitions has never been reported. Their most tangible fingerprint is the non-linear shape of the inverse susceptibility $1\!/\!\chi$ above 100\,K, i.e., in the temperature range $T\!>\!|J_i|$, where the Curie-Weiss regime is generally expected~\cite{millet2001,miller2012}. According to Pregelj~\textit{et al.}~\cite{pregelj2012}, the crystallographic symmetry of X=Br may be actually lower than $Pmmn$ considered so far (and also used in the present study), but no concluding information on the real symmetry is available. All these uncertainties hinder the quantitative microscopic analysis and restrict the scope of our model study to a qualitative level. 

The improvement of the microscopic magnetic model for francisites requires further experimental input, especially regarding the true crystallographic symmetry and the nature of the putative transitions at $100-150$\,K. Connections to other ferromagnetic kagome materials can be interesting as well. For example, kapellasite~\cite{fak2012} also exhibits ferromagnetic nearest-neighbor couplings on the kagome lattice and, additionally, frustrating second-neighbor AFM couplings. This compound lacks long-range magnetic order down to very low temperatures, which is reminiscent of the large classical degeneracy in the isotropic version of the francisite model. Although kapellasite features a different topology of AFM second-neighbor couplings $J_2$, francisite-like materials without the DM anisotropy will be definitely very interesting and have a high potential to exhibit fragile ordered states or even magnetic disorder.

\acknowledgments
We acknowledge fruitful discussions with Oleg Janson and Helge Rosner. AT was supported by the ESF through the Mobilitas grant MTT77 and by the IUT23-3 grant of the Estonian Research Agency. For the numerical CCM calculation we used the program package `The crystallographic CCM' by D. J. J. Farnell and J. Schulenburg, see \href{http://www-e.uni-magdeburg.de/jschulen/ccm/}{http://www-e.uni-magdeburg.de/jschulen/ccm/}.

\appendix


\section{Out-of-plane canting driven by $d_b'$ and $d_{1b}$}\label{app:oopansatz}
Here we provide the self-consistent equations that determine the parameters $\gamma_1$, $\gamma_2$, and $\theta$ of the non-coplanar ansatz of Eq.~(\ref{eq:oopansatz}). 
Experimentally~\cite{pregelj2012,miller2012}, the total moment of francisites points along the $\vec{c}$-axis, which indicates that the DM components along the $\vec{c}$ axis do not play any role. So we may replace $\phi=\pi/2$, and $d_{1c}=d_c'=d_{2c}=0$: 
\begin{widetext}
\bea\label{eq:Ecantedp}
\frac{\mc{E}}{N_{uc}S^2}&=& 
2J_1[ \cos(2\gamma_1)+\cos(2\gamma_3) ]
+4J_2 [\cos\gamma_1\cos\gamma_3\cos(2\theta)-\sin\gamma_1\sin\gamma_3]
+ 4 d_{2a} \cos\gamma_1\cos\gamma_3 \sin(2\theta)
\nonumber\\
&+&4(\cos\gamma_1+\cos\gamma_3) 
(J_1'\cos\theta - d_a' \sin\theta)
+4d_b' (\sin\gamma_1-\sin\gamma_3)
-2 d_{1b} \cos\theta \left[\sin(2\gamma_1) - \sin(2\gamma_3) \right]~,
\eea
where $N_{uc}=N/6$ is the number of unit cells. Minimizing with respect to $\theta$, $\gamma_1$ and $\gamma_3$, respectively, gives: 
\bea
&&
4\cos\gamma_1\cos\gamma_3[J_2\sin(2\theta)\!-\!d_{2a}\cos(2\theta)]
\!+\!2(\cos\gamma_1\!+\!\cos\gamma_3) 
(J_1'\sin\theta \!+\! d_a' \cos\theta)
\!-\!d_{1b} \sin\theta [\sin(2\gamma_1)\!-\!\sin(2\gamma_3)]
=0,~~~~~ \label{theta}\\
&&
J_1\sin(2\gamma_1)\!+\!J_2[\sin\gamma_1\cos\gamma_3\cos(2\theta)\!+\!\cos\gamma_1\sin\gamma_3]
\!+\!\sin\gamma_1(J_1'\cos\theta \!-\! d_a' \sin\theta)
\!-\!d_b'\cos\gamma_1\!+\!d_{1b}\cos\theta\cos(2\gamma_1)
\nonumber\\
&&~~+d_{2a}\sin\gamma_1\cos\gamma_3\sin(2\theta)
= 0~,~~~ \label{gamma1}\\
&&
J_1\sin(2\gamma_3)\!+\!J_2[\cos\gamma_1\sin\gamma_3\cos(2\theta)\!+\!\sin\gamma_1\cos\gamma_3]
\!+\!\sin\gamma_3(J_1'\cos\theta \!-\! d_a' \sin\theta)
\!+\!d_b'\cos\gamma_3\!-\!d_{1b}\cos\theta\cos(2\gamma_3)
\nonumber\\
&&~~+d_{2a}\cos\gamma_1\sin\gamma_3\sin(2\theta)
= 0~.~~~~~~~ \label{gamma3}
\eea
\end{widetext}
Note that (\ref{gamma1})-(\ref{gamma3}) map to each other under $\gamma_1\!\to\!-\gamma_3$, meaning that the solution satisfies the relation $\gamma_1\!=\!-\gamma_3$, in agreement with the insights based on Eq.~(\ref{eq:localBs}) above. Then, to leading order in $\gamma_{1,3}$, (\ref{theta}) and (\ref{gamma1}) give
\bea
&&
J_2 \sin(2\theta) \!+\! (J_1'\!-\!d_{1b}\gamma_1)\sin\theta \!+\! d_a' \cos\theta \!-\! d_{2a} \cos(2\theta)
\!=\!0, ~~~~~~\label{eq:theta3}\\
&&
\gamma_1 \!=\! \frac{d_b'-d_{1b}\cos\theta}{2J_1\!-\!2J_2 \sin^2\theta\!+\!J_1'\cos\theta
\!-\! d_a' \sin\theta\!+\!d_{2a}\sin(2\theta)}.~~~\label{eq:gamma13a}
\eea
Since $|d_{1b}\gamma_1|\ll |J_1'|$,  (\ref{eq:theta3}) reduces to (\ref{eq:theta2}), i.e., $\theta$ is unaffected by $d_b'$ and $d_{1b}$, to leading order. Finally, \ref{eq:gamma13a}) can be simplified to Eq.~(\ref{eq:gamma13}) of the main text.


\section{Low-field magnetization process for $\vec{H}\!\parallel\!\vec{c}$}\label{app:slopec}
Our aim here is to derive an expression for the slope of the low-field magnetization process along $\vec{c}$, above the metamagnetic transition field $H_{c,1}$. As shown in App.~\ref{app:oopansatz}, the angle $\theta$ is unaffected by the DM components $d_{1b}$ and $d_b'$ to lowest order, so we shall neglect these components here. We shall also neglect the anisotropy of the g-tensor as well as its inequivalence between Cu1 and Cu2 sites.

For $\vec{H}\!\parallel\!\vec{c}$, the classical GS is the canted phase of Eq.~(\ref{eq:canted}), but now the angle $\theta$ is not given by Eq.~(\ref{eq:theta2}), since it is now also influenced by the magnetic field. To avoid confusion we rename $\theta\!\to\!\theta_H$ and reserve the symbol $\theta$ for the zero-field value of the canting angle. Adding the Zeeman enegy $\mc{E}_Z$ in Eq.~(\ref{eq:Ecanted}) gives
\bea
&&\frac{\mc{E}+\mc{E}_Z}{N_{uc}} = 4 [J_1+J_2\cos(2\theta_H)+2J_1'\cos\theta_H-2d_a' \sin\theta_H\nonumber\\
&&~~+d_{2a}\sin(2\theta_H)] S^2-2g\mu_BH S \sin\phi(1+2\cos\theta_H)~.~~~
\eea
Minimizing with respect to $\phi$ and $\theta_H$ gives, respectively, $\phi=\pi/2$, and 
\be\label{eq:thetaB}
J_2\sin(2\theta_H)\!+\!(J_1'\!-\!\frac{g\mu_BH}{2S})\sin\theta_H\!+\!d_a'\cos\theta_H\!-\!d_{2a}\cos(2\theta_H)\!=\!0~,
\ee
which reduces to Eq.~(\ref{eq:theta2}) for $H=0$. Writing $\theta_H=\theta-\delta\theta$, and expanding (\ref{eq:thetaB}) to linear order in $\delta\theta$ and $g\mu_BH/J_1'$ gives
\be\label{eq:epsilon}
\delta\theta=\frac{g\mu_BH/(2S \sin\theta)}{2J_2+[d_a'-d_{2a}\cos\theta(1+2\sin^2\theta)]/\sin^3\theta}~,
\ee  
which in turn gives the slope $\kappa_c\!=\!\frac{M_c(H)-M_c(0)}{H}$ of the magnetization per site (for small fields):
\be
\kappa_c=\frac{2}{3}g\mu_BS\frac{\cos\theta_H-\cos\theta}{H} = \frac{2}{3}g\mu_B S\sin\theta \frac{\delta\theta}{H},
\ee 
which, in turn, leads to Eq.~(\ref{eq:slopec}) of the main text.

\section{Low-field magnetization process for $\vec{H}\!\parallel\!\vec{a}$}\label{app:slopea} 
Here we discuss the low-field magnetization process for $\vec{H}\!\parallel\!\vec{a}$. Again, we neglect the DM components $d_b'$ and $d_{1b}$.  Equivalently, we assume that the field is such that the out-of-plane canting driven by $d_b'$ and $d_{1b}$ is infinitesimal compared to that driven by the field. Starting from the zero-field GS of Eq.~(\ref{eq:canted}) with the global angle $\phi\!=\!\pi/2$, we can describe the field-induced out-of-plane canting by the classical ansatz given in Eq.~(\ref{eq:B//a}) of the main text. The total energy per unit cell, including the Zeeman energy $\mc{E}_Z$, is given by 
\begin{widetext}
\bea
\frac{\mc{E}\!+\!\mc{E}_Z}{N_{uc}}\!&=&\! 4\big[ J_1\!+\!J_2\left(\cos^2\beta_1\cos(2\theta_H)\!+\!\sin^2\beta_1\right)\!+\!2J_1'(\sin\beta_1\sin\beta_5\!+\!\cos\beta_1\cos\beta_5\cos\theta_H)\nonumber\\
&&
-2d_a'\cos\beta_1\cos\beta_5\sin\theta_H
+d_{2a} \cos^2\beta_1 \sin(2\theta_H) \big] S^2
-g\mu_BH S \left(4\sin\beta_1+2\sin\beta_5\right)~.
\eea
Minimizing with respect to $\theta_H$, $\beta_1$ and $\beta_5$, respectively, gives the set of equations
\bea
&&J_2 \cos\beta_1\sin(2\theta_H)+\cos\beta_5\left(J_1'\sin\theta_H+d_a'\cos\theta_H\right)-d_{2a}\cos\beta_1\cos(2\theta_H)
=0~,\label{eq:thetaB2}\\
&&J_2\sin(2\beta_1)\sin^2\theta_H+J_1' \left(\cos\beta_1\sin\beta_5-\sin\beta_1\cos\beta_5\cos\theta_H\right)\nonumber\\
&&~~~~+d_a'\sin\beta_1\cos\beta_5\sin\theta_H-\frac{1}{2}d_{2a}\sin(2\beta_1)\sin(2\theta_H)
=\frac{g\mu_BH}{2S}\cos\beta_1,~~~\label{eq:b1b5a}\\
&&J_1' \left(\sin\beta_1\cos\beta_5-\cos\beta_1\sin\beta_5\cos\theta_H\right) +d_a'\cos\beta_1\sin\beta_5\sin\theta_H
=\frac{g\mu_BH}{4S}\cos\beta_5.~~~\label{eq:b1b5b}
\eea
\end{widetext}
Writing again $\theta_H\!=\!\theta\!+\!\delta\theta$ and expanding to linear order in $\delta\theta$ as well as in $\beta_1$ and $\beta_5$, we find that (\ref{eq:thetaB2}) reduces to (\ref{eq:theta2}), i.e., the in-plane canting angle remains equal to the zero-field angle $\theta$ for small enough fields. Expanding the remaining two equations, (\ref{eq:b1b5a}) and (\ref{eq:b1b5b}), leads to two coupled equations for $\beta_1$ and $\beta_5$ which, in conjunction with (\ref{eq:theta2}), lead to: 
\bea
\beta_1&=& \cot\theta\frac{g\mu_BH}{2S}
\frac{J_1'(1+2\cos\theta)-2d_a'\sin\theta}{2d_a'(J_1'+d_{2a}\sin\theta)-2J_1' d_{2a} \cos\theta}, \nonumber\\
\beta_5 &=& \frac{g\mu_BH}{4S} 
\frac{J_1'(\csc\theta+2\cot\theta)+d_{2a}}{d_a'(J_1'+d_{2a}\sin\theta)-J_1'd_{2a}\cos\theta}.
\eea
The slope $\kappa_a$ of the magnetization per site for low fields is then given by $\kappa_a=M_a(H)/H=g\mu_B S (4\beta_1+2\beta_5)/(6H)$, which, in turn, leads to Eq.~(\ref{eq:slope-a}) of the main text.

\section{High-$T$ behavior of local susceptibilities for $\vec{H}\parallel\vec{a}$}\label{app:MF}
Here we discuss the influence of the DM anisotropy on the high-$T$ behavior of the local susceptibilities in the classical limit. To keep things as simple as possible we shall only include the DM components $d_a'$ and $d_{2a}$, and shall disregard the off-diagonal elements of the $\vec{g}$-tensors of the Cu1 sites. By symmetry then, the local moments satisfy the relations $\vec{m}_{1,2}\!\equiv\! \bs{\chi}_1\cdot\vec{H}$,  $\vec{m}_{3,4}\!\equiv\! \bs{\chi}_3\cdot\vec{H}$, and $\vec{m}_{5,6}\!\equiv\! \bs{\chi}_5\cdot\vec{H}$, where $\bs{\chi}_1$, $\bs{\chi}_3$, and $\bs{\chi}_5$ are the local susceptibility tensors. 
Let us suppose that the applied field is aligned with one of the crystallographic axes, $\vec{H}\parallel\vec{n}$. The single-site mean-field Hamiltonian reads
\be
\mc{H}_{\sf MF} = -g_n \mu_B \sum_{\vec{r}} \vec{H}\cdot \vec{S}_{\vec{r}},~~
\vec{H}_{\vec{r}}=\vec{H} - \frac{1}{g_n\mu_B}
\sum_{\vec{r}'} \bs{\Delta}_{\vec{r},\vec{r}'}\cdot\langle\vec{S}_{\vec{r}'}\rangle~,
\ee
where the second-rank tensors $\bs{\Delta}_{\vec{r},\vec{r}'}$ include the isotropic Heisenberg exchange as well as the anisotropic DM interactions. Within this mean-field approximation, the local magnetizations can be found self-consistently from Curie's law
$
\vec{m}_{\vec{r}} \!=\! g_n\mu_B\langle \vec{S}_{\vec{r}}\rangle \!=\! \frac{C}{T} \vec{H}_{\vec{r}}$, where $C\!=\!\frac{(g\mu_B)^2}{3k_B}S(S+1)$, and $S\!=\!1/2$. This gives:
\bea\label{eq:chis}
(t+2J_1)\vec{m}_1
+2(J_2-d_{2a}\vec{a}\times) \vec{m}_3
+2(J_1'+ d_a' \vec{a}\times) \vec{m}_5
=\vec{h}
\nonumber\\
2(J_2+d_{2a}\vec{a}\times) \vec{m}_1
+(t+2J_1)\vec{m}_3
+2(J_1'-d_a'\vec{a}\times) \vec{m}_5
=\vec{h}
\nonumber\\
2(J_1'-d_a'\vec{a}\times) \vec{m}_1
+2(J_1'+d_a'\vec{a}\times) \vec{m}_3
+t\vec{m}_5=\vec{h}
\nonumber
\eea
where $t\!\equiv\!(g\mu_B)^2T/C$ and $\vec{h}\!\equiv\!(g\mu_B)^2\vec{H}$.

Let us take for example the simplest possible case of $\vec{H}\!\parallel\!\vec{a}$. Here, by symmetry $\vec{m}_{1-4}\!=\!\chi_1^{aa} H\vec{a}$ and $\vec{m}_{5,6}\!=\!\chi_5^{aa} H\vec{a}$, and the above equations lead to
\bea
&&\chi_1^{aa} = (g_a\mu_B)^2\frac{(t-2J_1')}{t (t+2J_1+2J_2)-8J_1'^2},
\nonumber\\
&&\chi_5^{aa} = (g_a\mu_B)^2 \frac{(t+2J_1+2J_2-4J_1')}{t (t+2J_1+2J_2)-8J_1'^2},
\eea
and the total susceptibility per site $\chi^{aa}\!=\!\frac{1}{3}(2\chi_1^{aa}+\chi_5^{aa})$ is
\bea\label{eq:x1aa}
\chi^{aa} =
 (g_a\mu_B)^2 \frac{t+\frac{2}{3}(J_1+J_2)-\frac{8}{3}J_1'}{t (t+2J_1+2J_2)-8J_1'^2}~,
\eea
or equivalently
\bea\label{eq:1oxaa}
\frac{1}{\chi^{aa}} = \frac{T-\Theta_a}{C}-\frac{\gamma}{T-\Theta_a'}
\eea
where $\Theta_a=-(J_1+J_2+2J_1')/3$, $\Theta_a'=(4J_1'-J_1-J_2)/6$, and $\gamma C= \Theta_a \Theta_a' +J_1'^2/2$. While Eq.~(\ref{eq:1oxaa}) has the typical hyperbolic form for ferrimagnets~\cite{du2005magnetism}, it reduces to the well-known Curie-Weiss from for ferromagnets, by noting that $\Theta_a'$ turns out to be negative and of the order of $-20$\,K. As a result, the second term of Eq.~(\ref{eq:1oxaa}) is very small for $T\!>\!\Theta_a$ and can thus be disregarded for all practical purposes. We should also note that the response for $\vec{H}\parallel\vec{a}$ does not depend on the DM components $d_a'$ and $d_{2a}$.

\section{Linear spin-wave theory around any coplanar GS of the isotropic model}\label{app:lsw1}
{\it Hamiltonian and local frames} ---
Here we provide the details of the linear spin wave expansion around the coplanar states of the isotropic Hamiltonian. Using the Bravais lattice of the isotropic model (see Fig.~\ref{fig:model}), the total Hamiltonian can be written as 
\bea
\mc{H}_{\sf iso}&=&\sum_{\vec{L}}
J_1 \Big(
\vec{S}_{\vec{L},1}\cdot\vec{S}_{\vec{L},2}
+\vec{S}_{\vec{L}+\vec{t}_3,2}\cdot\vec{S}_{\vec{L}+\vec{t}_2,1}
\Big) \nonumber\\
&&
+J_1' \vec{S}_{\vec{L},3}\cdot \Big(
\vec{S}_{\vec{L},1}+
\vec{S}_{\vec{L},2}+
\vec{S}_{\vec{L}+\vec{t}_3,2}+
\vec{S}_{\vec{L}+\vec{t}_2,1}
\Big)\nonumber\\
&&
+J_2 
\Big( 
\vec{S}_{\vec{L},1}\cdot\vec{S}_{\vec{L}+\vec{t}_3,2}
+\vec{S}_{\vec{L},2}\cdot\vec{S}_{\vec{L}+\vec{t}_2,1}
\Big)~,
\eea
where $\vec{L}=n\vec{a}+m\vec{t}_2$ labels the position of the unit cell, $\vec{t}_2\!=\!(\vec{a}+\vec{b})/2$, and $\vec{t}_3=\vec{t}_2-\vec{a}$. 
Choosing the $\vec{xz}$-plane as the global spin plane of the coplanar states we write [see Eq.~(\ref{eq:cops})]: 
\bea \label{E2}
&&
\vec{S}_{\vec{L},\alpha}^{\text{clas}}=\cos\phi_{\alpha,m} \vec{z}+ \sin\phi_{\alpha,m}\vec{x},\nonumber\\
&&
\phi_{1,m}\!=\!\phi_{2,m}\!=\!\phi_0\!+\!2\theta\sum_{j=0}^{m-1}q_j, \nonumber\\
&&
\phi_{3,m}\!=\!\phi_{1,m}\!+\!q_m\theta~,~~ 
q_j\!=\!\pm1~,
\eea
where $\phi_0$ is arbitrary and the canting angle $\theta$ is given by Eq.~(\ref{eq:theta1}). 
Now, for each site of the lattice $(\vec{L},\alpha)\to\vec{r}$ we define a local coordinate system defined by three unit vectors $\vec{u}_{\vec{r}}$, $\vec{v}_{\vec{r}}\!=\!\vec{y}$, and $\vec{w}_{\vec{r}}=\vec{u}_{\vec{r}}\times\vec{v}_{\vec{r}}$, where the latter points along the direction of the spins in the classical state: 
\bea
\vec{w}_{\vec{r}} \!=\! \vec{S}_{\vec{r}}^{\text{clas}}/S,~
\vec{u}_{\vec{r}} \!=\! \vec{v}_{\vec{r}} \!\times\! \vec{w}_{\vec{r}} \!=\! \cos\phi_{\alpha,m}\vec{x}\!-\!\sin\theta_{\alpha,m}\vec{z}.~~~
\eea
Finally, we introduce the local exchange field exerted on site $\vec{r}$ via $\vec{B}_{\vec{r}}\!=\!-\sum_{\vec{r}'} J_{\vec{r},\vec{r}'}\vec{S}^{\text{clas}}_{\vec{r}'}$, which enters the classical GS energy   
$E_{\text{clas}} \!=\! \frac{1}{2}\sum_{\vec{r},\vec{r}'} J_{\vec{r},\vec{r}'}\vec{S}^{\text{clas}}_{\vec{r}} \cdot \vec{S}^{\text{clas}}_{\vec{r}'} 
\!=\! -\frac{S}{2}\sum_{\vec{r}} B_r$.

{\it Holstein-Primakoff transformation} ---
Next, we rewrite the spin operators as  
\bea
\vec{S}_{\vec{r}}&=&\left( \vec{S}_{\vec{r}}\cdot\vec{u}_{\vec{r}} \right) \vec{u}_{\vec{r}}
+\left( \vec{S}_{\vec{r}}\cdot\vec{v}_{\vec{r}} \right) \vec{v}_{\vec{r}}
+\left( \vec{S}_{\vec{r}}\cdot\vec{w}_{\vec{r}} \right) \vec{w}_{\vec{r}} \nonumber\\
&=& S_{\vec{r}}^u ~\vec{u}_{\vec{r}} + S_{\vec{r}}^v ~\vec{v}_{\vec{r}} + S_{\vec{r}}^w ~\vec{w}_{\vec{r}}~,
\eea
and we introduce bosonic operators $\{a_{\vec{r}}\}$ via the Holstein-Primakoff transformation: 
\be
S_{\vec{r}}^u \!\simeq\! \frac{\sqrt{S}}{\sqrt{2}} (a_{\vec{r}} \!+\! a_{\vec{r}}^+),~
S_{\vec{r}}^v \!\simeq\! -i \frac{\sqrt{S}}{\sqrt{2}} (a_{\vec{r}} \!-\! a_{\vec{r}}^+),~
S_{\vec{r}}^w \!=\! 
S\!-\!n_{\vec{r}},~
\ee
where $n_{\vec{r}} \!\equiv\! a_{\vec{r}}^+a_{\vec{r}}$. Any isotropic term in the Hamiltonian takes the following form 
\bea
&&\vec{S}_{\vec{r}}\!\cdot\!\vec{S}_{\vec{r}'} \!=\!
S^2 \vec{w}_{\vec{r}}\!\cdot\!\vec{w}_{\vec{r}'} \!+\!
\frac{S^{3/2}}{\sqrt{2}} \left[  \vec{u}_{\vec{r}}\!\cdot\!\vec{w}_{\vec{r}'} (a_{\vec{r}}\!+\!a_{\vec{r}}^+) 
\!+\! \vec{w}_{\vec{r}}\!\cdot\!\vec{u}_{\vec{r}'} (a_{\vec{r}'}\!+\!a_{\vec{r}'}^+)
\right] \nonumber\\
&&~~~+ 
\frac{S}{2} \Big[ 
\vec{u}_{\vec{r}}\!\cdot\!\vec{u}_{\vec{r}'}(a_{\vec{r}}\!+\!a_{\vec{r}}^+)( a_{\vec{r}'}\!+\!a_{\vec{r}'}^+ )
\!-\! \vec{v}_{\vec{r}}\!\cdot\!\vec{v}_{\vec{r}'} (a_{\vec{r}}\!-\!a_{\vec{r}}^+)(a_{\vec{r}'}\!-\!a_{\vec{r}'}^+) 
\nonumber\\
&&~~~
-2\vec{w}_{\vec{r}}\!\cdot\!\vec{w}_{\vec{r}'} (n_{\vec{r}}\!+\!n_{\vec{r}'})
\Big]
\!+\!\mc{O}(\sqrt{S})~.
\eea
The constant terms, which scale with $S^2$, give the classical GS energy $E_{\text{clas}}$. 
The one-magnon terms vanish since we expand around the classical minimum, and the quadratic terms, which scale with $S$, give:
\bea
&&
\mc{H}_{2-\text{magn}} \!=\!  \frac{1}{2}\frac{S}{2}\sum_{\vec{r},\vec{r}'}\Big[ 
J_{\vec{r},\vec{r}'}\vec{u}_{\vec{r}}\!\cdot\!\vec{u}_{\vec{r}'} (a_{\vec{r}}\!+\!a_{\vec{r}}^+)( a_{\vec{r}'}\!+\!a_{\vec{r}'}^+ )\nonumber\\
&&~~~
\!-\! J_{\vec{r},\vec{r}'}\vec{v}_{\vec{r}} \!\cdot\!\vec{v}_{\vec{r}'} (a_{\vec{r}}\!-\!a_{\vec{r}}^+)(a_{\vec{r}'}\!-\!a_{\vec{r}'}^+) 
\!-\!2J_{\vec{r},\vec{r}'}\vec{w}_{\vec{r}}\!\cdot\!\vec{w}_{\vec{r}'} (n_{\vec{r}}\!+\!n_{\vec{r}'})
\Big]~.\nonumber
\eea
Defining 
\be\label{eq:JuuJvv}
W_{\vec{r},\vec{r}'}^{(1,2)}\!\equiv\! J_{\vec{r},\vec{r}'}^{uu}\mp J_{\vec{r},\vec{r}'}^{vv}, 
J_{\vec{r},\vec{r}'}^{uu}\!\equiv\! J_{\vec{r},\vec{r}'} \vec{u}_{\vec{r}}\!\cdot\! \vec{u}_{\vec{r}'}, 
J_{\vec{r},\vec{r}'}^{vv}\!\equiv\! J_{\vec{r},\vec{r}'} \vec{v}_{\vec{r}}\!\cdot\! \vec{v}_{\vec{r}'}, 
\ee 
gives
\bea\label{eq:HTwoBoson}
\mc{H}_{\sf iso} &\simeq& 
S(S\!+\!1)\frac{E_{\text{clas}}}{S^2}
\!+\!\frac{S}{4}\sum_{\vec{r},\vec{r}'}\Big[ 
W^{(1)}_{\vec{r},\vec{r}'} a_{\vec{r}}a_{\vec{r}'} 
\!+\! W^{(2)}_{\vec{r},\vec{r}'} a_{\vec{r}}a_{\vec{r}'}^+ \!+\! h.c. \Big]\nonumber\\
&&~~
\!+\!\frac{1}{2} \sum_{\vec{r}} B_{\vec{r}} \left( a_{\vec{r}}^+ a_{\vec{r}}\!+\!a_{\vec{r}} a_{\vec{r}}^+\right) \!+\! \mc{O}(\sqrt{S})~.
\eea

Let us write down explicitly the parameters $B_{\vec{r}}$ and $W^{(1,2)}_{\vec{r},\vec{r}'}$ for all coplanar states. We find $\vec{B}_{\vec{L},\alpha}\!=\!B_ \alpha\vec{w}_{\vec{L}, \alpha}$, $\alpha \!=\!1$-$3$, with
\bea
&&B_1\!=\!B_2\!=\!\left\{\!\!
\begin{array}{ll}
-2S(J_1+J_2+J_1'), & \!\text{if}~0\!<\!J_2\!<\!|J_1'|/2\\
2S(J_2-J_1), &  \!\text{if}~J_2\!>\!|J_1'|/2 
\end{array}
\right.
, \nonumber\\
&&
B_3\!=\!-4SJ_1' \cos\theta\!=\!\left\{\!\!
\begin{array}{ll}
-4SJ_1', & \!\text{if}~0\!<\!J_2\!<\!|J_1'|/2\\
+2SJ_1'^2/J_2, & \!\text{if}~J_2\!>\!|J_1'|/2
\end{array}
\right. , \nonumber\\
&&
W^{(1,2)}_{\vec{L},1,\vec{L}+\vec{t}_3,2}=W^{(1,2)}_{\vec{L},2,\vec{L}+\vec{t}_2,1}=J_2\left(\cos(2\theta) \mp 1\right)~,\nonumber\\
&&
W^{(1,2)}_{\vec{L},3,\vec{L},1}=W^{(1,2)}_{\vec{L},3,\vec{L},2}=W^{(1,2)}_{\vec{L},3,\vec{L}+\vec{t}_2,1}=W^{(1,2)}_{\vec{L},3,\vec{L}+\vec{t}_3,2}\nonumber\\
&&~~~~~~=J_1'\left(\cos\theta \mp 1\right)~, \nonumber\\
&&
W^{(1,2)}_{\vec{L},1,\vec{L},2}=W^{(1,2)}_{\vec{L}+\vec{t}_3,2,\vec{L}+\vec{t}_2,1}=(1\mp 1)J_1~.\nonumber
\eea
At this point we see that none of the 
parameters that enter in (\ref{eq:HTwoBoson}) depend on the constants $q_m$ and $\phi_0$
[see Eq.~(\ref{E2})], i.e., the quadratic spin wave spectrum and the corresponding zero-point energies are identical for all coplanar states.

{\it Final quadratic form in momentum space} --- 
Using $a_{\vec{k},j}  = \frac{1}{\sqrt{N/3}}\sum_{\vec{L}} e^{i \vec{k}\cdot \vec{L}} a_{\vec{L},j}$, and defining 
\bea
&&
f(\vec{k})\!=\!J_2\cos^2\theta (e^{i\vec{k}\cdot\vec{t}_3}+e^{-i\vec{k}\cdot\vec{t}_2})+J_1(1+e^{-i\vec{k}\cdot\vec{t}_1}),\nonumber\\
&&
g(\vec{k})\!=\!-J_2\sin^2\theta (e^{i\vec{k}\cdot\vec{t}_3}\!+\!e^{-i\vec{k}\cdot\vec{t}_2})~, \nonumber\\
&&
h(\vec{k})\!=\!-J_1'\sin^2\frac{\theta}{2}(1\!+\!e^{i\vec{k}\cdot\vec{t}_2}), 
p(\vec{k})\!=\!-J_1'\sin^2\frac{\theta}{2}(1\!+\!e^{i\vec{k}\cdot\vec{t}_3})~, \nonumber\\
&&
q(\vec{k})\!=\!J_1'\cos^2\frac{\theta}{2}(1\!+\!e^{i\vec{k}\cdot\vec{t}_2}),
r(\vec{k})\!=\!J_1'\cos^2\frac{\theta}{2}(1\!+\!e^{i\vec{k}\cdot\vec{t}_3})~,\nonumber
\eea
we may group all terms in a matrix form 
\be\label{eq:compact}
\mc{H}_{\sf iso} \!=\! S(S\!+\!1)\frac{E_{\text{class}}}{S^2} \!+\! \frac{1}{2} \sum_{\vec{k}} \vec{A}_{\vec{k}}^+ \!\cdot\! \vec{M}_{\vec{k}} \!\cdot\! \vec{A}_{\vec{k}}\!+\!\mc{O}(\sqrt{S})~,
\ee
with $\vec{A}_{\vec{k}}^+ \!=\! \left( a_{\vec{k},1}^+, a_{\vec{k},2}^+,  a_{\vec{k},3}^+, a_{-\vec{k},1}, a_{-\vec{k},2}, a_{-\vec{k},3} \right)$, and
\small
\be
\vec{M}_{\vec{k}} \!=\! S\left(\begin{array}{c|c|c|c|c|c}
B_1/S&f(\vec{k})&q(-\vec{k})&0&g(\vec{k})&h(-\vec{k})\\
&&&&&\\
f(-\vec{k})&B_1/S&r(-\vec{k})&g(-\vec{k})&0&p(-\vec{k})\\
&&&&&\\
q(\vec{k})&r(\vec{k})&B_3/S&h(\vec{k})&p(\vec{k})&0\\
&&&&&\\
0&g(\vec{k})&h(-\vec{k})&B_1/S&f(\vec{k})&q(-\vec{k})\\
&&&&&\\
g(-\vec{k})&0&p(-\vec{k})&f(-\vec{k})&B_1/S&r(-\vec{k})\\
&&&&&\\
h(\vec{k})&p(\vec{k})&0&q(\vec{k})&r(\vec{k})&B_3/S\\
\end{array}
\right).~~~
\ee
\normalsize
We next define the commutator matrix
$
\vec{g} \!=\! \vec{A}_{\vec{k}} \!\cdot\!\vec{A}_{\vec{k}}^\dagger - \left( (\vec{A}_{\vec{k}}^\dagger)^T \!\cdot\! \vec{A}_{\vec{k}}^T \right)^T
=\left(
\begin{array}{c|c}
1\!\text{l}_3 & 0 \\
\hline
0 & -1\!\text{l}_3 
\end{array}
\right)
$,  
where $1\!\text{l}_3$ stands for the identity $3\!\times\!3$ matrix, and perform a Bogoliubov transformation~\cite{Blaizot,White} $\vec{A}_{\vec{k}}\!=\!\vec{S}_{\vec{k}}\!\cdot\!\tilde{\vec{A}}_{\vec{k}}$, which must conserve the commutation relations $\tilde{\vec{g}}\!=\!\vec{g}$, and at the same time diagonalize the Hamiltonian, namely
\bea
\mc{H}_{\sf iso} \!&=&\! S(S\!+\!1) \frac{E_{\text{class}}}{S^2}  \!+\! \frac{1}{2} \sum_{\vec{k}} \tilde{\vec{A}}_{\vec{k}}^\dagger \!\cdot\! (\vec{S}^\dagger_{\vec{k}} \vec{M}_{\vec{k}} \vec{S}_{\vec{k}}) \!\cdot\! \tilde{\vec{A}}_{\vec{k}}\!+\!\mc{O}(\sqrt{S}) \nonumber\\
&=&
S(S\!+\!1)\frac{E_{\text{class}}}{S^2} \!+\! \frac{1}{2} \sum_{\vec{k}}  \tilde{\vec{A}}_{\vec{k}}^\dagger \!\cdot\! \vec{\Omega}_{M_{\vec{k}}} \!\cdot\! \tilde{\vec{A}}_{\vec{k}}\!+\!\mc{O}(\sqrt{S}),
\eea
where $\vec{\Omega}_{M_{\vec{k}}}$ is diagonal and can be found from the eigenvalue equation $(\vec{g} \vec{M}_{\vec{k}}) \cdot \vec{S}_{\vec{k}} = \vec{S}_{\vec{k}} \cdot (\vec{g} \vec{\Omega}_{M_k}) \equiv \vec{S}_{\vec{k}} \cdot \vec{\Omega}_{g M_{\vec{k}}}$. Denoting the eigenvalues of $\vec{\Omega}_{M_{\vec{k}}}$ by $\omega_{1-3}(\vec{k})$, we finally get
\be\label{eq:finaldiagform1}
\mc{H}_{\sf iso} = S(S+1) \frac{E_{\text{class}}}{S^2}  +  \sum_{k,\alpha} \omega_\alpha(\vec{k}) \left( \tilde{a}_{\vec{k},\alpha}^+ \tilde{a}_{\vec{k},\alpha} +\frac{1}{2}\right) + \mc{O}(\sqrt{S}) ~.
\ee
The dispersions of the three spin-wave branches are shown in Fig.~\ref{fig:LSWSpectra}(a). 
The GS of the quadratic Hamiltonian is the vacuum of the bosonic operators $\tilde{a}_{k,1}$, $ \tilde{a}_{k,2}$, $\tilde{a}_{k,3}$, and so the quadratic zero-point energy correction is given by $\delta E^{(2)} = \frac{1}{2} \sum_{\vec{k},\alpha} \omega_\alpha(\vec{k})$, and can be found by a numerical integration over the BZ of the isotropic model. The results are shown in Fig.~\ref{fig:CCM} (b). As expected, there are no quantum corrections for $J_2\leq 0.5 |J_1|$, where the GS is the fully polarized state, while for $J_2\!>\!0.5 |J_1|$, the zero-point energy increases by increasing the frustrating coupling $J_2$.


\section{Linear spin-wave theory around the canted phase in the presence of DM anisotropies}\label{app:lsw2}
Here we include the DM anisotropy on all bonds, and perform the corresponding semiclassical expansion around the canted phase of (\ref{eq:oopansatz}), where the parameters $\gamma_1\!=\!-\gamma_3$ and $\theta$ are given by Eq.~(\ref{theta}-\ref{gamma1}). 
To this end we must now use the actual symmetry of the crystal and denote the spin sites by $\vec{r}\to(\vec{R},\alpha)$, with $\alpha=1$-$6$ (see Fig.~\ref{fig:model}). This means that we now have six bosons per unit cell. 
The local coordinate frames are chosen so that $\vec{w}_{\vec{R},\alpha}\!=\!\vec{S}_{\vec{R},\alpha}/S$, as given by (\ref{eq:oopansatz}), 
$\vec{v}_{\vec{R},\alpha}\!=\!\vec{a}$, and $\vec{u}_{\vec{R},\alpha}\!=\!\vec{a}\!\times\!\vec{w}_{\vec{R},\alpha}$.

Due to the DM anisotropy, the interaction terms in the spin Hamiltonian are of the form $\vec{S}_{\vec{r}}\!\cdot\!\bs{\Gamma}_{\vec{r},\vec{r}'}\!\cdot\!\vec{S}_{\vec{r}'}$, where $\bs{\Gamma}_{\vec{r},\vec{r}'}$ is a tensor. For the isotropic exchange couplings $\Gamma_{\vec{r},\vec{r}'}^{\alpha\beta}\!=\!J_{\vec{r},\vec{r}'}\delta_{\alpha\beta}$, while for the DM couplings $\Gamma_{\vec{r},\vec{r}'}^{\alpha\beta}\!=\!\epsilon_{\alpha\beta \gamma} D_{\vec{r},\vec{r}'}^\gamma$.
Following the same steps as above, the quadratic bosonic Hamiltonian takes the form (\ref{eq:HTwoBoson}), but now the coefficients 
$W^{(1,2)}_{\vec{r},\vec{r}'}$ generalize to 
\bea
&&W^{(1,2)}_{\vec{r},\vec{r}'}\equiv \Gamma_{\vec{r},\vec{r}'}^{uu}
\mp i \Gamma_{\vec{r},\vec{r}'}^{uv}
-i\Gamma_{\vec{r},\vec{r}'}^{vu}
\mp \Gamma_{\vec{r},\vec{r}'}^{vv},\nonumber\\
&&
\Gamma^{uu}_{\vec{r},\vec{r}'}\equiv\vec{u}_{\vec{r}}\cdot\bs{\Gamma}_{\vec{r},\vec{r}'}\cdot\vec{u}_{\vec{r}'},~~~
\Gamma^{uv}_{\vec{r},\vec{r}'}\equiv\vec{u}_{\vec{r}}\cdot\bs{\Gamma}_{\vec{r},\vec{r}'}\cdot\vec{v}_{\vec{r}'}, \text{etc}.~~~
\eea
Compared to (\ref{eq:JuuJvv}), we now have finite contributions from the coefficients $\Gamma^{uv}_{\vec{r},\vec{r}'}$ and $\Gamma^{vu}_{\vec{r},\vec{r}'}$, because the local axes $\vec{u}_{\vec{r}}$ are not any longer perpendicular to $\vec{v}_{\vec{r}}\!=\!\vec{a}$, due to the out-of-plane canting which, in turn, is generated by $d_b'$ and $d_{1b}$.
The parameters $B_{\vec{R},j}=B_j$ and $W^\pm_{\vec{R},j,\vec{R}+\bs{\gamma},j'}$ are given by:
\begin{widetext}
\bea
&&
B_{1-4}/(2S)\!=\!\sin\gamma_1 [-d_b' \!+\! (J_1 \!-\! J_2)\sin\gamma_1] 
\!+\! \cos\gamma_1 [-\cos\theta (J_1' \!-\! 2 d_{1b} \sin\gamma_1) \!+\! d_a' \sin\theta] 
\!-\! \cos^2\gamma_1 [J_1 \!+\! J_2 \cos(2\theta) \!+\! d_{2a} \sin(2\theta)], \nonumber\\
&& 
B_{5-6}/(4S) \!=\! (-J_1' \cos\theta \!+\! d_a' \sin\theta) \cos\gamma_1 \!-\!d_b' \sin\gamma_1, \nonumber\\
&&
W^{(1)}_{\vec{R},1,\vec{R},2}=W^{(1)}_{\vec{R},3,\vec{R},4} = -J_1 \sin^2\gamma_1 \equiv 2 q_1 , \nonumber\\
&&
W^{(2)}_{\vec{R},1,\vec{R},2}=W^{(2)}_{\vec{R},3,\vec{R},4}=J_1(1+\cos^2\gamma_1)+2 i d_{1b}\cos\gamma_1\sin\theta \equiv 2 q_2,\nonumber\\
&&
W^{(1)}_{\vec{R},1,\vec{R}-\vec{a},4}=W^{(1)}_{\vec{R},2,\vec{R},3}=-J_2+\cos^2\gamma_1[J_2\cos(2\theta)+d_{2a}\sin(2\theta)] \equiv 2 f_1 ,\nonumber\\
&&
W^{(2)}_{\vec{R},1,\vec{R}-\vec{a},4}=W^{(2)}_{\vec{R},2,\vec{R},3}=J_2+\cos^2\gamma_1[J_2\cos(2\theta)+d_{2a}\sin(2\theta] \equiv 2 f_2 ,\nonumber\\
&&
W^{(1)}_{\vec{R},5,\vec{R},1}=W^{(1)}_{\vec{R},5,\vec{R},3}=-J_1'+\cos\gamma_1[J_1'\cos\theta-d_a'\sin\theta+i d_b' \sin\theta] \equiv 2g_1,\nonumber\\
&&
W^{(2)}_{\vec{R},5,\vec{R},1}=W^{(2)}_{\vec{R},5,\vec{R},3}=J_1'+\cos\gamma_1[J_1'\cos\theta-d_a'\sin\theta+i d_b' \sin\theta] \equiv 2g_2,\nonumber\\
&&
W^{(1)}_{\vec{R},5,\vec{R},2}=W^{(1)}_{\vec{R},5,\vec{R},4}%
= 2g_1^\ast,~~
W^{(2)}_{\vec{R},5,\vec{R},2}=W^{(2)}_{\vec{R},5,\vec{R},4}%
= 2g_2^\ast.~~\nonumber
\eea
\end{widetext} 
With $a_{\vec{k},\alpha}=\frac{1}{\sqrt{N/6}}\sum_{\vec{R}} e^{i \vec{k}\cdot \vec{R}} a_{\vec{R},\alpha}$, and collecting all terms leads again to the compact matrix form of (\ref{eq:compact}), where
\begin{widetext}
\bea
\vec{A}_\vec{k}^+ = \left(
\begin{array}{cccccccccccccc}
a_{\vec{k},1}^+, & a_{\vec{k},2}^+, & a_{\vec{k},3}^+, & a_{\vec{k},4}^+, & a_{\vec{k},5}^+, & a_{\vec{k},6}^+, & a_{-\vec{k},1}, & a_{-\vec{k},2}, & a_{-\vec{k},3}, & a_{-\vec{k},4}, & a_{-\vec{k},5}, & a_{-\vec{k},6} 
\end{array}
\right), \nonumber
\eea
\small
\be
\frac{\vec{M}_k}{S}\!\!=\!\!\left(\!\!
\begin{array}{c|c|c|c|c|c|c|c|c|c|c|c}
B_1/S&q_2 l_a^\ast&0&h_2^\ast e^{-ik_a}&g_2^\ast&g_2^\ast e^{-ik_{ab}}&0&q_1 l_a^\ast &0&h_1^\ast e^{-ik_a}&g_1&g_1 e^{-ik_{ab}}\\
&&&&&&&&&&&\\
q_2^\ast l_a&B_1/S&h_2(-\vec{k})&0&g_2&g_2e^{-ik_b}&q_1 l_a&0&h_1^\ast &0&g_1^\ast&g_1^\ast e^{-ik_b}\\
&&&&&&&&&&&\\
0&h_2 &B_1/S&q_2 l_a^\ast &g_2^\ast&g_2^\ast&0&h_1 &0&q_1 l_a^\ast&g_1&g_1\\
&&&&&&&&&&&\\
h_2 e^{ik_a}&0&q_2^\ast l_a &B_1/S&g_2e^{ik_a}&g_2&h_1 e^{ik_a}&0&q_1 l_a&0&g_1^\ast e^{ik_a}&g_1^\ast \\
&&&&&&&&&&&\\
g_2 &g_2^\ast&g_2&g_2^\ast e^{-ik_a}&B_5/S&0&g_1&g_1^\ast&g_1&g_1^\ast e^{-ik_a}&0&0\\
&&&&&&&&&&&\\
g_2 e^{ik_{ab}}&g_2^\ast e^{ik_b}&g_2&g_2^\ast&0&B_5/S&g_1e^{ik_{ab}}&g_1^\ast e^{ik_b}&g_1&g_1^\ast&0&0\\
&&&&&&&&&&&\\
0&q_1 l_a^\ast&0&h_1^\ast e^{-ik_a}&g_1^\ast&g_1^\ast e^{-ik_{ab}}&B_1/S&q_2^\ast l_a^\ast&0&h_2^\ast e^{-ik_a}&g_2&g_2e^{-ik_{ab}}\\
&&&&&&&&&&&\\
q_1 l_a&0&h_1^\ast &0&g_1 &g_1 e^{-ik_b}&q_2 l_a&B_1/S&h_2^\ast &0&g_2^\ast&g_2^\ast e^{-ik_b}\\
&&&&&&&&&&&\\
0&h_1&0&q_1 l_a^\ast &g_1^\ast&g_1^\ast&0&h_2 &B_1/S&q_2^\ast l_a^\ast &g_2&g_2\\
&&&&&&&&&&&\\
h_1 e^{ik_a}&0&q_1 l_a &0&g_1 e^{ik_a}&g_1 &h_2 e^{ik_a}&0&q_2 l_a &B_1/S&g_2^\ast e^{ik_a}&g_2^\ast\\
&&&&&&&&&&&\\
g_1^\ast&g_1&g_1^\ast&g_1 e^{-ik_a}&0&0&g_2^\ast&g_2&g_2^\ast&g_2e^{-ik_a}&B_5/S&0\\
&&&&&&&&&&&\\
g_1^\ast e^{ik_{ab}}&g_1 e^{ik_b}&g_1^\ast&g_1&0&0&g_2^\ast e^{ik_{ab}}&g_2 e^{ik_b}&g_2^\ast&g_2&0&B_5/S\\
\end{array}
\right)\nonumber
\ee
\end{widetext}
\normalsize
where we have defined $k_a\equiv\vec{k}\cdot\vec{a}$, $k_b\equiv\vec{k}\cdot\vec{b}$, $k_{ab}\!\equiv\!k_a+k_b$, $l_a\!\equiv\!1+e^{i k_a}$, and $h_{1,2}(\vec{k})\equiv f_{1,2} (1+e^{ik_b})$. 
A Bogoliubov transformation leads again to the diagonal form (\ref{eq:finaldiagform1}), with the difference that we now have six spin-wave branches instead of three.
The dispersions of these branches are shown in Fig.~\ref{fig:LSWSpectra}(b) for $d_a'\!=\!d_{2a}\!=\!0$ and in Fig.~\ref{fig:LSWSpectra}(c) for $d_a'\!=\!0.15|J_1|$ and $d_{2a}\!=\!0$. 

\subsection{Special directions in momentum space due to glide planes}\label{app:glides}
As we discussed in the main text, in the presence of all DM couplings the GS preserves the glide plane operations that involve non-primitive translations along the diagonal directions of the lattice, $(\vec{a}\pm\vec{b})/2$, followed by a reflection in the $ab$-plane. Here we shall discuss the consequences of this symmetry in the excitation spectrum along the special lines $\vec{k}\!=\!(k_a,\pi)$ and $(\pi,k_b)$.
Let us denote by $\mc{G}$ the glide plane involving the translation by $(-\vec{a}+\vec{b})/2$. Applying for example this operation to the spin operator $\vec{S}_{\vec{R},1}$ gives 
\bea
&&\mc{G}\cdot \vec{S}_{\vec{R},1} =  -S_{\vec{R}-\vec{a},3}^a\vec{a} -S_{\vec{R}-\vec{a},3}^b\vec{b} + S_{\vec{R}-\vec{a},3}^c\vec{c} \nonumber\\ 
\Rightarrow &&~
\mc{G} \cdot \vec{S}_{\vec{k},1} = e^{-i\vec{k}\cdot\vec{a}}\,\left( -S_{\vec{k},3}^a\vec{a} -S_{\vec{k},3}^b\vec{b} + S_{\vec{k},3}^c\vec{c}\right)~.~~
\eea
Using the Holstein-Primakoff transformation and the relations between the local axes $\vec{w}_{\vec{R},\alpha}$ for different $\alpha$, we arrive at
\bea\label{eq:glides}
&&\mc{G}\cdot a_{\vec{k},1}^+ = - e^{-i\vec{k}\cdot\vec{a}}\,a_{\vec{k},3}^+,~~~\mc{G}\cdot a_{\vec{k},2}^+ = - e^{-i\vec{k}\cdot\vec{a}}\,a_{\vec{k},4}^+, \nonumber\\
&&\mc{G}\cdot a_{\vec{k},3}^+ = - e^{+i\vec{k}\cdot\vec{b}}\,a_{\vec{k},1}^+,~~~\mc{G}\cdot a_{\vec{k},4}^+ = - e^{+i\vec{k}\cdot\vec{b}}\,a_{\vec{k},2}^+,\\
&&\mc{G}\cdot a_{\vec{k},5}^+ = - e^{-i\vec{k}\cdot\vec{a}}\,a_{\vec{k},6}^+,~~~\mc{G}\cdot a_{\vec{k},6}^+ = - e^{+i\vec{k}\cdot\vec{b}}\,a_{\vec{k},5}^+. \nonumber
\eea
The matrix $M_{\vec{k}}$ is invariant under this transformation for the special lines $\vec{k}\!=\!(k_a,\pi)$ and $(\pi,k_b)$, which explains why the six spin-wave branches organize into three doubly-degenerate branches along these special lines.

\subsection{Field dependence for $\vec{H}\parallel\vec{c}$}
For comparison to experiments, we also examine the influence of a magnetic field $H$ along the $\vec{c}$-axis. To keep things simple, we shall disregard the weak DM components along the $\vec{b}$-axis, $d_b'$ and $d_{1b}$, and we shall also assume that the $\vec{c}$-axis is a principal axis for the Cu1 sites too, with spectroscopic factor $g_c$. The linear spin-wave theory proceeds as above, but i) we must replace everywhere $\theta\to\theta_H$, because we expand around the classical canted state of App.~\ref{app:slopec}, and ii) we must add the contribution from the Zeeman energy:
\bea
&&\mc{H}_Z = -g_c\mu_B H \sum_{\vec{R},\alpha} \vec{S}_{\vec{R},\alpha}\cdot\vec{c}\nonumber\\
&&~~~
=-g\mu_B H \sum_{\vec{R},\alpha}\left(S_{\vec{R},\alpha}^u \vec{u}_{\vec{R},\alpha}\cdot\vec{c}+S_{\vec{R},\alpha}^w \vec{w}_{\vec{R},\alpha}\cdot\vec{c}\right).~
\eea
Collecting the quadratic terms from the Holstein-Primakoff transformation gives, in momentum space:
\bea
&&\mc{H}_Z^{(2)} \!=\! g_c\mu_B H \sum_{\vec{k}}\! \Big\{\left(n_{\vec{k},1}\!+\!n_{\vec{k},2}\!+\!n_{\vec{k},3}\!+\!n_{\vec{k},4}\right) \cos\theta_H \nonumber\\
&&~~~\!+\! \left(n_{\vec{k},5}\!+\!n_{\vec{k},6}\right) \Big\}.
\eea
So the final form of the quadratic boson Hamiltonian is the same as before but with the replacements $B_1\!\to\!B_1\!+\!g_c\mu_BH\cos\theta_H$ and $B_5\!\to\!B_5\!+\!g_c\mu_BH$.

\subsection{Excitation energies at the $\bs{\Gamma}$-point with $\vec{H}\parallel\vec{c}$}\label{app:Gmodes}
In the following we shall extract a number of analytical expressions for the special case of $d_b'\!=\!d_{1b}\!=\!0$, in the presence of a field $H$ along the $\vec{c}$-axis. We first discuss the modes at the $\bs{\Gamma}$-point. As discussed in the main text, the isotropic spectrum shown in Fig.~\ref{fig:LSWSpectra}(b) features two Golstone modes, on account of the higher symmetry in the absence of any anisotropy. The spectra of Fig.~\ref{fig:LSWSpectra}(c) show that a finite $d_a'$ gives rise to a finite spin-gap for one of the two Goldstone modes. It turns out that the eigenvector and the corresponding eigenvalue $\Delta_{\Gamma,2}$ of $\vec{g}\vec{M}$ corresponding to this $\vec{k}\!=\!0$ mode is given by 
\begin{widetext}
\bea
&&
\vec{v}_{\Gamma,2}=\frac{1}{2} \Big(\cos{u_2},\cos{u_2},-\cos{u_2},-\cos{u_2},0,0,-\sin{u_2},-\sin{u_2},\sin{u_2},\sin{u_2},0,0\Big)~,\nonumber\\
&&
\Delta_{\Gamma,2} = 2S \Big[ (1-\tan{u_2}) \sin\theta_H (J_2\sin\theta_H-d_{2a}\cos\theta_H) + d_a'\csc\theta_H - d_{2a}\cot\theta_H \Big] + g_c\mu_B \cos\theta_H H~,\label{eqapp:Delta2}\\
&&
\sin(2u_2)=\frac{\sin\theta_H (J_2\sin\theta_H-d_{2a}\cos\theta_H)}{J_2\sin^2\theta_H+d_a' \csc\theta_H-d_{2a}\cos\theta_H(\csc\theta_H+\sin\theta_H)+ g_c\mu_B H \cos\theta_H}~.\nonumber
\eea
\end{widetext}
This mode describes an AFM canting mode of the Cu1 spins, where $\vec{S}_{\vec{R},1}$ and $\vec{S}_{\vec{R},3}$ rotate in phase with $\vec{S}_{\vec{R},2}$ and $\vec{S}_{\vec{R},4}$, respectively, $\vec{S}_{\vec{R},1-2}$ rotate in opposite directions from $\vec{S}_{\vec{R},3-4}$, while the directions of the Cu2 spins remain unaffected. This explains why $\Delta_{\Gamma,2}$ does not depend on $J_1$.

Turning to the third mode, its energy $\Delta_{\Gamma,3}$ is given in Eq.~(\ref{eq:Delta3}) and reads
\be
\vec{v}_{\Gamma,3}=(0,0,0,0,\frac{1}{\sqrt{2}},-\frac{1}{\sqrt{2}},0,0,0,0,0,0)~, 
\ee
which describes an out-of-phase rotation of the two Cu2 sites, without rotating the Cu1 sites. This explains why $\Delta_{\Gamma,3}$ does not depend on the Cu1-Cu1 couplings $J_2$, $J_1$, and $d_{2a}$.

For the fourth and sixth modes, the eigenvectors and eigenenergies read
\begin{widetext}
\bea
&&\vec{v}_{\Gamma,4}=\frac{1}{2}\left(-\cos{u_4},\cos{u_4},-\cos{u_4},\cos{u_4},0,0,\sin{u_4},-\sin{u_4},\sin{u_4},-\sin{u_4},0,0\right)~,\nonumber\\
&&
\Delta_{\Gamma,4}=2S \Big[
-2J_1+(1-\tan{u_4}) \sin\theta_H ( J_2\sin\theta_H-d_{2a}\cos\theta_H )+d_a'\csc\theta_H-d_{2a} \cot\theta_H \Big] + g_c\mu_B \cos\theta_H H~,\nonumber\\
&&
\sin(2u_4)=\frac{\sin\theta_H(J_2 \sin\theta_H-d_{2a}\cos\theta_H)}{-2J_1+J_2 \sin^2\theta_H+d_a'\csc\theta_H-d_{2a} \cos\theta_H (\sin\theta_H+\csc\theta_H)+ g_c\mu_B H
\cos\theta_H}
\nonumber\\
&&\vec{v}_{\Gamma,6}=\frac{1}{2}\left(\cos{u_6},-\cos{u_6},-\cos{u_6},\cos{u_6},0,0,\sin{u_6},-\sin{u_6},-\sin{u_6},\sin{u_6},0,0\right)~,\nonumber\\
&&
\Delta_{\Gamma,6}=2S\Big[
-2J_1+2J_2-(1+\tan{u_6}) \sin\theta_H (J_2\sin\theta_H-d_{2a}\cos\theta_H)  -d_{2a}\cot\theta
+ d_a'\csc\theta_H \Big]+g_c\mu_B \cos\theta_H H
~,\nonumber\\
&&
\sin(2u_6)=\frac{\sin\theta_H (J_2\sin\theta_H-d_{2a}\cos\theta_H)}{-2J_1+J_2(1+\cos^2\theta_H) + d_a' \csc\theta_H  + d_{2a}\cos\theta_H (\sin\theta_H-\csc\theta_H) +g_c \mu_B \cos\theta_H H}~.
\eea
\end{widetext}
The fourth mode describes an out-of-plane canting of the Cu1 spins, $\vec{S}_{\vec{R},1-3}$ rotate in phase with $\vec{S}_{\vec{R},2-4}$, while the directions of the Cu2 spins remain unaffected. This mode is then energetically favored by $d_b'$ and $d_{1b}$, because they give rise to the same relative signs in the out-of-plane canting angles. 
The sixth mode is similar to $\vec{v}_{\Gamma,4}$, but now $\vec{S}_{\vec{R},1-4}$ rotate in phase with $\vec{S}_{\vec{R},2-3}$, which is energetically favored by $d_c'$ and $d_{1c}$.

\subsection{Lowest excitation gap at the $\vec{k}\!=\!(0,\pi)$ point}
As shown in Fig.~\ref{fig:LSWSpectra}\,(c), at $\vec{k}\!=\!(0,\pi)$ (point A in Fig.~\ref{fig:LSWSpectra}\,d) there are two low-lying modes that are degenerate, due to the glide planes discussed  above. Their excitation energy, $\Delta_{\text{A},1}$, vanishes in the isotropic limit $d_{a}'\!=\!d_{2a}\!=\!0$, as shown in Fig.~\ref{fig:LSWSpectra}\,(b). The two modes correspond to the eigenvectors (related to each other via the glide transformation of Eq.~(\ref{eq:glides})):
\bea
&&\vec{v}_{\text{A},1}=(z_1, z_1, 0, 0, z_5, -z_5, -z_7, -z_7, 0, 0, -z_{11},  z_{11}),\nonumber\\
&&\vec{v}_{\text{A},2}=(0, 0, -z_1, -z_1, -z_5, -z_5, 0, 0, z_7, z_7, z_{11}, z_{11}),\nonumber
\eea
where the constants $z_1$, $z_5$, $z_7$ and $z_{11}$ can be numerically determined in terms of the microscopic parameters of the model. The gap $\Delta_{\text{A},1}$ can be found using the following equations:
\be\label{eqapp:DeltaA1}
\Delta_{\text{A},1}^2=2S \left(c_2-\sqrt{c_2^2+16 c_0}\right)/8,
\ee
where 
\begin{widetext}
\bea
&&c_0 \frac{\sin^2(2\theta_H)}{J_1' \cos\theta_H- d_a' \sin\theta_H} =
J_1' \Big\{
4 d_{2a} d_a' (7  + \cos(4\theta_H) )
- d_a'^2 [\cos(3\theta_H) +\cos(5\theta_H) +14 \cos\theta_H]
- 16 d_{2a}^2 \cos\theta_H
\Big\}\nonumber\\
&&~~~
+J_1'^2 \Big\{
4 d_a'[3\sin\theta_H-\sin(3\theta_H)]
+2d_{2a}[\sin(4\theta_H)-2\sin(2\theta_H)]
\Big\}\nonumber\\
&&~~~
+ d_a'^3 [2\sin\theta_H+3\sin(3 \theta_H) +\sin(5 \theta_H)]
- 2 d_a'^2 d_{2a}  [6\sin(2\theta_H) +\sin(4 \theta_H)]
+  16 d_a' d_{2a}^2 \sin\theta_H,~~~\nonumber\\
&&
c_2=J_1'^2 (16 \cos^2\theta_H+8\cos\theta_H+\sec^2\theta_H) 
- 4J_1' d_a' [\csc(2\theta_H)+2\sin\theta_H+4\sin(2\theta_H)]
+ J_1' d_{2a} [4\sec\theta_H\csc(2\theta_H)]\nonumber\\
&&
+ d_a'^2 (16\sin^2\theta_H + \csc^2\theta_H )
-4 d_a' d_{2a} \csc\theta_H \csc(2\theta_H) 
+ 4 d_{2a}^2 \csc^2(2\theta_H)~.\nonumber
\eea
\end{widetext}
In the limit of $d_{2a}=0$ and $d_a'\ll J_1'$, $\Delta_{\text{A},1}$ is given by
\be\label{eq:DeltaA1}
\Delta_{\text{A},1}\simeq 4S \frac{\sqrt{- J_1'd_a' \tan\theta_H}}{\sqrt{8\cos\theta_H+16\cos^2\theta_H+\sec^2\theta_H}}~.
\ee
In analogy to $\Delta_{\Gamma,2}$, this mode scales with the geometric mean of the DM anisotropy and the corresponding exchange energy.


%

\end{document}